\documentclass[11pt,a4paper]{article}
\usepackage{psfrag}
\usepackage{graphics}
\usepackage{graphicx}
\usepackage{float}
\usepackage{amssymb,epsfig,amsmath,euscript,array}
\usepackage{cite}
\usepackage{multicol}
\usepackage{subfig}
\usepackage{fancyhdr}
\usepackage{color}
\usepackage{pdflscape}
\usepackage{hyperref}
\usepackage{graphicx}
\usepackage{tikz}
\usetikzlibrary{arrows.meta}

\makeatletter
\@addtoreset{equation}{section}
\makeatother

\usepackage[us,24hr]{datetime}

\newcounter{multieqs}

\newcommand{\be}{\begin{equation}}
\newcommand{\ee}{\end{equation}}

\newcommand{\bm}[1]{\mbox{\boldmath $#1$}}

\def\bd{\begin{document}}
\def\ed{\end{document}}
\def\nn{\nonumber}
\def\bea{\begin{eqnarray}}
\def\eea{\end{eqnarray}}
\let\bm=\bibitem
\let\la=\label

\def\npb#1#2#3{Nucl. Phys. {\bf{B#1}} #3 (#2)}
\def\plb#1#2#3{Phys. Lett. {\bf{#1B}} #3 (#2)}
\def\prl#1#2#3{Phys. Rev. Lett. {\bf{#1}} #3 (#2)}
\def\prd#1#2#3{Phys. Rev. {D \bf{#1}} #3 (#2)}
\def\cmp#1#2#3{Comm. Math. Phys. {\bf{#1}} #3 (#2)}
\def\cqg#1#2#3{Class. Quantum Grav. {\bf{#1}} #3 (#2)}
\def\nppsa#1#2#3{Nucl. Phys. B (Proc. Suppl.) {\bf{#1A}}#3 (#2)}
\def\ap#1#2#3{Ann. of Phys. {\bf{#1}} #3 (#2)}
\def\ijmp#1#2#3{Int. J. Mod. Phys. {\bf{A#1}} #3 (#2)}
\def\rmp#1#2#3{Rev. Mod. Phys. {\bf{#1}} #3 (#2)}
\def\mpla#1#2#3{Mod. Phys. Lett. {\bf A#1} #3 (#2)}
\def\jhep#1#2#3{J. High Energy Phys. {\bf #1} #3 (#2)}
\def\atmp#1#2#3{Adv. Theor. Math. Phys. {\bf #1} #3 (#2)}

\newcommand{\EQ}[1]{\begin{equation} #1 \end{equation}}
\newcommand{\AL}[1]{\begin{subequations}\begin{align} #1 \end{align}\end{subequations}}
\newcommand{\SP}[1]{\begin{equation}\begin{split} #1 \end{split}\end{equation}}
\newcommand{\ALAT}[2]{\begin{subequations}\begin{alignat}{#1} #2 \end{alignat}\end{subequations}}
\def\beqa{\begin{eqnarray}}
\def\eeqa{\end{eqnarray}}
\def\beq{\begin{equation}}
\def\eeq{\end{equation}}

\def\N{{\cal N}}
\def\sst{\scriptscriptstyle}
\def\thetabar{\bar\theta}
\def\Tr{{\rm Tr}}
\def\one{\mbox{1 \kern-.59em {\rm l}}}
 \def\Nh{\hat{N}}

\def\a{\alpha}      \def\da{{\dot\alpha}}
\def\b{\beta}       \def\db{{\dot\beta}}
\def\c{\gamma}  \def\G{\Gamma}  \def\cdt{\dot\gamma}
\def\d{\delta}  \def\D{\Delta}  \def\ddt{\dot\delta}
\def\e{\epsilon}        \def\vare{\varepsilon}
\def\f{\phi}    \def\F{\Phi}    \def\vvf{\f}
\def\h{\eta}
\def\k{\kappa}
\def\l{\lambda} \def\L{\Lambda}
\def\m{\mu} \def\n{\nu}
\def\o{\omega}
\def\p{\pi} \def\P{\Pi}
\def\s{\sigma}  \def\S{\Sigma}
\def\t{\tau}
\def\th{\theta} \def\Th{\Theta} \def\vth{\vartheta}
\def\X{\Xeta}
\def\z{\zeta}

\def\cA{{\cal A}} \def\cB{{\cal B}} \def\cC{{\cal C}}
\def\cD{{\cal D}} \def\cE{{\cal E}} \def\cF{{\cal F}}
\def\cG{{\cal G}} \def\cH{{\cal H}} \def\cI{{\cal I}}
\def\cJ{{\cal J}} \def\cK{{\cal K}} \def\cL{{\cal L}}
\def\cM{{\cal M}} \def\cN{{\cal N}} \def\cO{{\cal O}}
\def\cP{{\cal P}} \def\cQ{{\cal Q}} \def\cR{{\cal R}}
\def\cS{{\cal S}} \def\cT{{\cal T}} \def\cU{{\cal U}}
\def\cV{{\cal V}} \def\cW{{\cal W}} \def\cX{{\cal X}}
\def\cY{{\cal Y}} \def\cZ{{\cal Z}}

\def\ua{\underline{\alpha}}
\def\ub{\underline{\phantom{\alpha}}\!\!\!\beta}
\def\uc{\underline{\phantom{\alpha}}\!\!\!\gamma}
\def\um{\underline{\mu}}
\def\ud{\underline\delta}
\def\ue{\underline\epsilon}
\def\una{\underline a}\def\unA{\underline A}
\def\unb{\underline b}\def\unB{\underline B}
\def\unc{\underline c}\def\unC{\underline C}
\def\und{\underline d}\def\unD{\underline D}
\def\une{\underline e}\def\unE{\underline E}
\def\unf{\underline{\phantom{e}}\!\!\!\! f}\def\unF{\underline F}
\def\unm{\underline m}\def\unM{\underline M}
\def\unn{\underline n}\def\unN{\underline N}
\def\unp{\underline{\phantom{a}}\!\!\! p}\def\unP{\underline P}
\def\unq{\underline{\phantom{a}}\!\!\! q}
\def\unQ{\underline{\phantom{A}}\!\!\!\! Q}
\def\unH{\underline{H}}

\def\As {{A \hspace{-6.4pt} \slash}\;}
\def\bs {{b \hspace{-6.4pt} \slash}\;}
\def\Ds {{D \hspace{-6.4pt} \slash}\;}
\def\ds {{\del \hspace{-6.4pt} \slash}\;}
\def\ss {{\s \hspace{-6.4pt} \slash}\;}
\def\ks {{ k \hspace{-6.4pt} \slash}\;}
\def\ps {{p \hspace{-6.4pt} \slash}\;}
\def\pas {{{p_1} \hspace{-6.4pt} \slash}\;}
\def\pbs {{{p_2} \hspace{-6.4pt} \slash}\;}

\def\Fh{\hat{F}}
\def\Vh{\hat{V}}
\def\Xh{\hat{X}}
\def\ah{\hat{a}}
\def\xh{\hat{x}}
\def\yh{\hat{y}}
\def\ph{\hat{p}}
\def\xih{\hat{\xi}}

\def\psit{\tilde{\psi}}
\def\Psit{\tilde{\Psi}}
\def\tht{\tilde{\th}}
\def\lt{\tilde{\lambda}}
\def\At{\tilde{A}}
\def\Qt{\tilde{Q}}
\def\Rt{\tilde{R}}
\def\Nt{\tilde{N}}

\def\at{\tilde{a}}
\def\st{\tilde{s}}
\def\ft{\tilde{f}}
\def\pt{\tilde{p}}
\def\qt{\tilde{q}}
\def\vt{\tilde{v}}
\def\nt{\tilde{n}}

\def\delb{\bar{\partial}}
\def\bz{\bar{z}}
\def\bD{\bar{D}}
\def\bB{\bar{B}}

\def\bk{{\bf k}}
\def\bl{{\bf l}}
\def\bp{{\bf p}}
\def\bq{{\bf q}}
\def\br{{\bf r}}
\def\bx{{\bf x}}
\def\by{{\bf y}}
\def\bR{{\bf R}}
\def\bV{{\bf V}}

\def\d{\delta}\def\D{\Delta}\def\ddt{\dot\delta}

\def\pa{\partial} \def\del{\partial}
\def\xx{\times}
\def\uno{\mbox{1 \kern-.59em {\rm l}}}

\def\trp{^{\top}}
\def\inv{^{-1}}
\def\dag{{^{\dagger}}}
\def\pr{^{\prime}}
\def\lan{\langle}
\def\ran{\rangle}

\def\rar{\rightarrow}
\def\lar{\leftarrow}
\def\lrar{\leftrightarrow}

\newcommand{\0}{\,\!}      
\def\one{1\!\!1\,\,}
\def\im{\imath}
\def\jm{\jmath}

\newcommand{\tr}{\mbox{tr}}
\newcommand{\slsh}[1]{/ \!\!\!\! #1}

\def\vac{|0\rangle}
\def\lvac{\langle 0|}

\def\hlf{\frac{1}{2}}
\def\ove#1{\frac{1}{#1}}

\def\Box{\square}
\def\ZZ{\mathbb{Z}}
\def\CC#1{({\bf #1})}
\def\bcomment#1{}
\def\bfhat#1{{\bf \hat{#1}}}
\def\VEV#1{\left\langle #1\right\rangle}

\newcommand{\ex}[1]{{\rm e}^{#1}} \def\ii{{\rm i}}

\def\rr{{\rm r}} \def\rs{{\rm s}}\def\rv{{\rm v}}
\def\ri{{\rm i}}\def\rj{{\rm j}}

\newcommand{\lrbrk}[1]{\left(#1\right)}
\newcommand{\sfrac}[2]{{\textstyle\frac{#1}{#2}}}

\font\mybb=msbm10 at 12pt
\def\bb#1{\hbox{\mybb#1}}

\font\myBB=msbm10 at 18pt
\def\BB#1{\hbox{\myBB#1}}

\newcommand{\tclr}{\textcolor}
\newcommand{\bpmat}{\begin{pmatrix}}
\newcommand{\epmat}{\end{pmatrix}}
\newcommand{\mrm}[1]{\mathrm{#1}}
\newcommand{\mrs}[1]{\scriptscriptstyle{\mathrm{#1}}}
\newcommand{\vct}[1]{\boldsymbol{#1}}
\newcommand{\hf}{\frac{1}{2}}
\newcommand{\x}{\times}
\newcommand{\pd}{\partial}
\newcommand{\dslash}{\displaystyle{\not}}
\newcommand{\ol}[1]{\overline{#1}}
\newcommand{\abs}[1]{\vert{#1}\vert}
\newcommand{\chiSqM}{\chi^2_{\mrm{min}}}
\newcommand{\chiSqMDof}{\chi^2_{\mrm{min}}/\mrm{d.o.f.}}
\newcommand{\om}{\omega}
\newcommand{\Lag}{\mathcal{L}}
\newcommand{\ord}{\mathcal{O}}
\newcommand{\eps}{\epsilon}
\newcommand{\beFrac}{\frac{1-\be}{1+\be}}
\newcommand{\beFracI}{\frac{1+\be}{1-\be}}
\newcommand{\amu}{a_{\mu}}
\newcommand{\damu}{\delta\amu}
\newcommand{\Damu}{\Delta\amu}
\newcommand{\amuUnit}{10^{-10}}
\newcommand{\mmu}{m_{\mu}}
\newcommand{\amuQED}{\amu^{\mrm{QED}}}
\newcommand{\amuEW}{\amu^{\mrm{EW}}}
\newcommand{\amuEWl}{\amu^{\mrm{EW,}\,1l}}
\newcommand{\amuEWll}{\amu^{\mrm{EW,}\,2l}}
\newcommand{\amuh}{\amu^{\mrm{had}}}
\newcommand{\amuhLO}{\amu^{\text{had, LOVP}}}
\newcommand{\amuhHO}{\amu^{\text{had, HOVP}}}
\newcommand{\amuhHOa}{\amu^{\text{had, HOVP(a)}}}
\newcommand{\amuhHOb}{\amu^{\text{had, HOVP(b)}}}
\newcommand{\amuhHOc}{\amu^{\text{had, HOVP(c)}}}
\newcommand{\amuhLbL}{\amu^{\text{had, LbL}}}
\newcommand{\ff}[3]{\mathcal{F}_{\pi^{0{#1}}\gamma^{#2}\gamma^{#3}}}
\newcommand{\alps}{\alpha_s}
\newcommand{\asmz}{\alpha_s(M_Z^2)}
\newcommand{\amz}{\alpha(M_Z^2)}
\newcommand{\aqmz}{\alpha_{\mrm{QED}}(M_Z^2)}
\newcommand{\delAlp}{\Delta\alpha}
\newcommand{\dAlpL}{\delAlp_{\mrm{lep}}}
\newcommand{\dAlpT}{\delAlp_{\mrm{top}}}
\newcommand{\dAlpH}{\delAlp_{\mrm{had}}}
\newcommand{\dAlpHF}{\dAlpH^{(5)}}
\newcommand{\dAlpHFmz}{\dAlpHF(M_Z^2)}
\newcommand{\tmin}{t_{\mrm{min}}}
\newcommand{\sTh}{s_{\mrm{th}}}
\newcommand{\eTh}{\sqrt{\sTh}}
\newcommand{\Ekmi}{E^{\,(k,m)}_i}
\newcommand{\Nkm}{N^{(k,m)}}
\newcommand{\Nkn}{N^{(k,n)}}
\newcommand{\Nexp}{N_{\mrm{exp}}}
\newcommand{\Nclu}{N_{\mrm{clu}}}
\newcommand{\Ntot}{N_{\mrm{tot}}}
\newcommand{\Rkmi}{R^{\,(k,m)}_i}
\newcommand{\Rknj}{R^{\,(k,n)}_j}
\newcommand{\dRkmi}{\mrm{d}\Rkmi}
\newcommand{\dRtkmi}{\mrm{d}\tilde{R}^{\,(k,m)}_i}
\newcommand{\BR}[2]{\mathcal{B}(#1\to #2)}
\newcommand{\decay}[2]{#1\to #2}
\newcommand{\UpsIVs}{\Upsilon(4S)}
\newcommand{\Gee}{\Gamma_{ee}}
\newcommand{\Gtot}{\Gamma_{\mrm{tot}}}
\newcommand{\ppC}{\pi^+\pi^-}
\newcommand{\ppN}{\pi^0\pi^0}
\newcommand{\pppC}{\pi^+\pi^-\pi^0}
\newcommand{\kkC}{K^+K^-}
\newcommand{\kskl}{K^0_S K^0_L}
\newcommand{\ksks}{K^0_S K^0_S}
\newcommand{\klkl}{K^0_L K^0_L}
\newcommand{\kskp}{K^0_S K^{\pm}\pi^{\mp}}
\newcommand{\eeMuMu}{e^+e^-\to\mu^+\mu^-}
\newcommand{\eeHadr}{e^+e^-\to\mrm{hadrons}}
\newcommand{\eeGhadr}{e^+e^-\to\gamma^*\to\mrm{hadrons}}
\newcommand{\tauNuHadr}{\tau\to\nu_{\tau}+\mrm{hadrons}}
\newcommand{\eeGPiPi}{e^+e^-\to\gamma^*\to\pi^+\pi^-}
\newcommand{\tauNuWNuPiPi}{\tau\to\nu_{\tau}W\to\nu_{\tau}\pi\pi^0}
\newcommand{\eeGIncl}{e^+e^-\to\gamma^*\to\mrm{all\,hadrons}}
\newcommand{\eeIncl}{e^+e^-\to\mrm{all\,hadrons}}
\newcommand{\eePiG}{e^+e^-\to\pi^0\gamma}
\newcommand{\eePiPi}{e^+e^-\to\pi^+\pi^-}
\newcommand{\eePiPiPi}{e^+e^-\to\pi^+\pi^-\pi^0}
\newcommand{\eeKK}{e^+e^-\to K^+K^-}
\newcommand{\ch}{\mrm{ch}}
\newcommand{\iso}{\mrm{iso}}
\newcommand{\noeta}{\text{no }\eta}
\newcommand{\kkr}{K\bar{K}\rho}
\newcommand{\kkp}{K\bar{K}\pi}
\newcommand{\kkpp}{K\bar{K}2\pi}
\newcommand{\kkppp}{K\bar{K}3\pi}
\newcommand{\isoAA}{(2\pi^+2\pi^-\pi^0)_{\mrm{no}\,\eta}}
\newcommand{\isoAB}{(\pi^+\pi^-3\pi^0)_{\mrm{no}\,\eta}}
\newcommand{\isoAC}{\omega(\to\mrm{npp})2\pi}
\newcommand{\isoACf}{\omega(\to\text{non-pure pionic states})2\pi}
\newcommand{\isoAD}{\eta\pi^+\pi^-}
\newcommand{\isoBA}{(2\pi^+2\pi^-2\pi^0)_{\mrm{no}\,\eta}}
\newcommand{\isoBB}{(\pi^+\pi^-4\pi^0)_{\mrm{no}\,\eta}}
\newcommand{\isoBC}{3\pi^+3\pi^-}
\newcommand{\isoBD}{\omega(\to\mrm{npp})3\pi}
\newcommand{\isoBDf}{\omega(\to\text{non-pure pionic state})3\pi}
\newcommand{\isoBE}{\eta\omega}
\newcommand{\isoEA}{\kkppp}
\newcommand{\isoEAa}{(K^+K^-\pi^+\pi^-\pi^0)_{\mrm{no}\,\eta}}
\newcommand{\isoEAb}{(K^0\bar{K}^0\pi^+\pi^-\pi^0)_{\mrm{no}\,\eta}}
\newcommand{\isoEB}{\omega(\to\mrm{npp})K\bar{K}}
\newcommand{\isoEBf}{\omega(\to\text{non-pure pionic states})K\bar{K}}
\newcommand{\isoEC}{\eta\phi}
\newcommand{\isoFA}{\eta2\pi^+2\pi^-}
\newcommand{\isoFB}{\eta\pi^+\pi^-2\pi^0}
\newcommand{\sigEEhadr}{\sigma(\eeHadr)}
\newcommand{\sigHad}{\sigma_{\mrm{had}}}
\newcommand{\sigHadB}{\sigHad^0}
\newcommand{\sigPt}{\sigma_{\mrm{pt}}}
\newcommand{\Rhad}{R_{\mrm{had}}}
\newcommand{\affinfnuniA}[3]{Dipartimento di Fisica dell'Universit\`a #1\\and INFN Sezione di #2, #3, Italy}
\newcommand{\affinfnuniB}[3]{Dipartimento di Scienze Fisiche dell'Universit\`a #1\\and INFN Sezione di #2, #3, Italy}
\newcommand{\affinfnuniC}[3]{Dipartimento di Fisica dell'Universit\`a #1\\and INFN gruppo collegato di #2, #3, Italy}
\newcommand{\affuni}[2]{Dipartimento di Fisica dell'Universit\`a #1, #2, Italy.}
\newcommand{\affinfn}[2]{INFN Sezione di #1, #2, Italy.}
\newcommand{\newauthor}[2]{\rule{0pt}{1.1\baselineskip}#2$^{#1}$}

\newcommand\blfootnote[1]{%
  \let\thempfn\relax
  \footnotetext[0]{\emph{#1}}
}

\begin{document}

\setcounter{page}{1}
\thispagestyle{empty}
\begin{flushright}
LTH 1162 
\end{flushright}

\begin{center}

\hspace{150cm}
\
\\
\
\\
\
\\
\
\\
\
\\
\
\\
{\Large {\bf Combination of KLOE $\sigma\big(e^+e^-\rightarrow\pi^+\pi^-\gamma(\gamma)\big)$ measurements and determination of $a_{\mu}^{\pi^+\pi^-}$ in the energy range $0.10 < s < 0.95$ GeV$^2$ } \\}
\
\\
\
\
\\
  {\Large The KLOE-2 Collaboration\\}
  \vspace{0.2cm}
  {\large \newauthor{f,d}{A.~Anastasi},
\newauthor{d}{D.~Babusci},
\newauthor{d,w}{M.~Berlowski},
\newauthor{d}{C.~Bloise},
\newauthor{d}{F.~Bossi},
\newauthor{t}{P.~Branchini},
\newauthor{s,t}{A.~Budano},
\newauthor{v}{L.~Caldeira~Balkest\r{a}hl},
\newauthor{v}{B.~Cao},
\newauthor{s,t}{F.~Ceradini},
\newauthor{d}{P.~Ciambrone},
\newauthor{d}{F.~Curciarello},
\newauthor{c}{E.~Czerwi\'nski},
\newauthor{o,p}{G.~D'Agostini},
\newauthor{d}{E.~Dan\`e},
\newauthor{r}{V.~De~Leo},
\newauthor{d}{E.~De~Lucia},
\newauthor{d}{A.~De~Santis},
\newauthor{d}{P.~De~Simone},
\newauthor{s,t}{A.~Di~Cicco},
\newauthor{o,p}{A.~Di~Domenico},
\newauthor{d}{D.~Domenici},
\newauthor{d}{A.~D'Uffizi},
\newauthor{q,r}{A.~Fantini},
\newauthor{e}{G.~Fantini},
\newauthor{d}{P.~Fermani},
\newauthor{u,p}{S.~Fiore},
\newauthor{c}{A.~Gajos},
\newauthor{o,p}{P.~Gauzzi},
\newauthor{d}{S.~Giovannella},
\newauthor{t}{E.~Graziani},
\newauthor{h,i}{V.~L.~Ivanov},
\newauthor{v}{T.~Johansson},
\newauthor{c}{D.~Kisielewska-Kami\'nska},
\newauthor{d}{X.~Kang},
\newauthor{h,i}{E.~A.~Kozyrev},
\newauthor{w}{W.~Krzemien},
\newauthor{v}{A.~Kupsc},
\newauthor{s,t}{S.~Loffredo},
\newauthor{h,i}{P.~A.~Lukin},
\newauthor{g,b}{G.~Mandaglio},
\newauthor{d,n}{M.~Martini},
\newauthor{q,r}{R.~Messi},
\newauthor{d}{S.~Miscetti},
\newauthor{d}{G.~Morello},
\newauthor{r}{D.~Moricciani},
\newauthor{c}{P.~Moskal},
\newauthor{t}{A.~Passeri},
\newauthor{m,p}{V.~Patera},
\newauthor{d}{E.~Perez~del~Rio},
\newauthor{r}{N.~Raha},
\newauthor{d}{P.~Santangelo},
\newauthor{k,l}{M.~Schioppa},
\newauthor{s,t}{A.~Selce},
\newauthor{c}{M.~Silarski},
\newauthor{d}{F.~Sirghi},
\newauthor{h,i}{E.~P.~Solodov},
\newauthor{t}{L.~Tortora},
\newauthor{j,1}{G.~Venanzoni}\blfootnote{$^1$ Corresponding authors: \texttt{graziano.venanzoni@lnf.infn.it, a.i.Keshavarzi@liverpool.ac.uk}},%
\newauthor{w}{W.~Wi\'slicki},
\newauthor{v}{M.~Wolke}
}
\
\\
\
\\
and
\
\\
\
\\
  {\large A. Keshavarzi$^{x,}$\footnotemark[1], S. E. M\"{u}ller$^y$ and T. Teubner$^x$}
\\
\
\\
  {\small \em \center $^a$\affinfn{Bari}{Bari}}
  {\small \em \center $^b$\affinfn{Catania}{Catania}}
  {\small \em \center $^c$Institute of Physics, Jagiellonian University, Cracow, Poland.}
  {\small \em \center $^d$Laboratori Nazionali di Frascati dell'INFN, Frascati, Italy.}
  {\small \em \center $^e$Gran Sasso Science Institute, L'Aquila, Italy.}
  {\small \em \center $^f$Dipartimento di Scienze Matematiche e Informatiche, Scienze Fisiche e Scienze della Terra dell'Universit\`a di Messina, Messina, Italy.}
  {\small \em \center $^g$Dipartimento di Scienze Chimiche, Biologiche, Farmaceutiche ed Ambientali dell'Universit\`a di Messina, Messina, Italy.}
  {\small \em \center $^h$Budker Institute of Nuclear Physics, Novosibirsk, Russia.}
  {\small \em \center $^i$Novosibirsk State University, Novosibirsk, Russia.}
  {\small \em \center $^j$\affinfn{Pisa}{Pisa}}
  {\small \em \center $^k$\affuni{della Calabria}{Rende}}
  {\small \em \center $^l$INFN Gruppo collegato di Cosenza, Rende, Italy.}
  {\small \em \center $^m$Dipartimento di Scienze di Base ed Applicate per l'Ingegneria dell'Universit\`a ``Sapienza'', Roma, Italy.}
  {\small \em \center $^n$Dipartimento di Scienze e Tecnologie applicate, Universit\`a ``Guglielmo Marconi", Roma, Italy.}
  {\small \em \center $^o$\affuni{``Sapienza''}{Roma}}
  {\small \em \center $^p$\affinfn{Roma}{Roma}}
  {\small \em \center $^q$\affuni{``Tor Vergata''}{Roma}}
  {\small \em \center $^r$\affinfn{Roma Tor Vergata}{Roma}}
  {\small \em \center $^s$Dipartimento di Matematica e Fisica dell'Universit\`a 
``Roma Tre'', Roma, Italy.}
  {\small \em \center $^t$\affinfn{Roma Tre}{Roma}}
  {\small \em \center $^u$ENEA, Department of Fusion and Technology for Nuclear Safety and Security, Frascati (RM), Italy}
  {\small \em \center $^v$Department of Physics and Astronomy, Uppsala University, Uppsala, Sweden.}
  {\small \em \center $^w$National Centre for Nuclear Research, Warsaw, Poland.}
    
{\small \em \center $^x$Department of Mathematical Sciences, University of Liverpool, Liverpool L69 3BX, UK.}
{\small \em \center $^y$Department of Information Services and Computing \& Institute of Radiation Physics, Helmholtz-Zentrum Dresden-Rossendorf, Dresden, Germany.} 
\
\\
\
\\
\
\\
{\small \center {\bf  Data vectors and covariance matrices are available at} \\ \texttt{\url{http://www.lnf.infn.it/kloe/ppg/ppg_2017/ppg_2017.html}}}  \\

\
\\
\
\\
{\normalsize \bf Abstract}
\end{center}
The three precision measurements of the cross section $\sigma\big(e^+e^-\rightarrow\pi^+\pi^-\gamma(\gamma)\big)$  using initial state radiation  by the KLOE collaboration provide an important input for the prediction of the hadronic contribution to the anomalous magnetic moment of the muon. These measurements are correlated for both statistical and systematic uncertainties and, therefore, the simultaneous use of these measurements requires covariance matrices that fully describe the correlations. We present the construction of these covariance matrices and use them to determine a combined KLOE measurement for $\sigma\big(e^+e^-\rightarrow\pi^+\pi^-\gamma(\gamma)\big)$. We find, from this combination, a two-pion contribution to the muon magnetic anomaly in the energy range $0.10 < s < 0.95$ GeV$^2$ of $a_{\mu}^{\pi^+\pi^-} = (489.8 \pm 1.7_{\rm stat} \pm 4.8_{\rm sys} ) \times 10^{-10}$.



\newpage


\section{Introduction}

The KLOE collaboration have made three precise measurements of the cross section $\allowbreak\sigma\big(e^+e^-\allowbreak\rightarrow\pi^+\pi^-\gamma(\gamma)\big)$ in 2008~\cite{Ambrosino:2008aa,KLOE08-KLOEnote}, 2010~\cite{Ambrosino:2010bv,KLOE10-KLOEnote} and 2012~\cite{Babusci:2012rp,KLOE12-KLOEnote}.\footnote{The KLOE collaboration also made a measurement of $\sigma\big(e^+e^-\rightarrow\pi^+\pi^-\gamma(\gamma)\big)$ in 2005~\cite{Aloisio:2004bu}. However, this is now considered to be superseded by the 2008 measurement, as discussed in~\cite{Ambrosino:2008aa}.} These measurements are crucial for estimating the hadronic vacuum polarisation (HVP) contribution to the anomalous magnetic moment of the muon, $a_{\mu}^{\rm HVP}$, which is presently the limiting factor in the precision of the Standard Model (SM) prediction, $a_{\mu}^{\rm SM}$. This SM prediction disagrees with the experimental value, $a_{\mu}^{\rm exp}$~\cite{Bennett:2002jb,Bennett:2004pv,Bennett:2006fi,PDG2016}, by approximately $3.5$ standard deviations or higher~\cite{Keshavarzi:2018mgv,Hagiwara:2011af,Davier:2010nc,Davier:2017zfy,Jegerlehner:2009ry,Jegerlehner:2017lbd,Benayoun:2015gxa,Ananthanarayan:2016mns}, making it an interesting probe of potential physics beyond the SM. Currently, the uncertainties of $a_{\mu}^{\rm SM}$ and $a_{\mu}^{\rm exp}$ are of comparable magnitude. However, with new experimental efforts at Fermilab~\cite{Grange:2015fou} and J-PARC~\cite{Mibe:2010zz} set to improve the experimental error by a factor of four compared to the BNL measurements~\cite{Bennett:2002jb,Bennett:2004pv,Bennett:2006fi}, it is imperative that the SM prediction is also improved.

The HVP contribution to  $a_{\mu}^{\rm SM}$ can be determined using a dispersion integral and the cross section $\sigma^0\big(e^+e^-\rightarrow{\rm hadrons}\big)$, which is bare (undressed of all vacuum polarisation (VP) effects) as indicated by the superscript `$0$', but includes final state radiation (FSR). At leading order (LO), the dispersion integral is
\beq \label{eq:amu}
a_{\mu}^{\rm LO,HVP} = \frac{1}{4\pi^3}\int^{\infty}_{s_{th}} {\rm d}s \ \sigma^0_{\rm had}(s)K(s) \ ,
\eeq
where $s_{th} = m_{\pi^0}^2$ is the hadronic production threshold, $\sigma^0_{\rm had}(s)$ is the bare cross section of the process $e^+e^-\rightarrow{\rm hadrons}$ and $K(s)$ is a well-known kernel function~\cite{Brodsky:1967sr,Lautrup:1969fr}. The contribution of the $\pi^+\pi^-$ final state to the anomalous magnetic moment of the muon, $a_{\mu}^{\pi^+\pi^-}$, is over 70\% of the total estimate of $a_{\mu}^{\rm HVP}$ and is also the largest contributor to its uncertainty. Consequently, the three measurements of the cross section $\sigma^0\big(e^+e^-\rightarrow\pi^+\pi^-\gamma(\gamma)\big)$ by the KLOE collaboration are invaluable to precisely determine $a_{\mu}^{\pi^+\pi^-}$.

The simultaneous input of the KLOE measurements into equation~\eqref{eq:amu} requires a detailed analysis to attain the correct combination of the three, which will have a non-trivial influence on $a_{\mu}^{\pi^+\pi^-}$ and provide an important comparison with other experimental measurements of $\sigma_{\pi\pi}$. The KLOE measurements of $\sigma_{\pi\pi(\gamma)}$ are, in part, highly correlated, necessitating the construction of full statistical and systematic covariance matrices to be used in any combination of these data. To combine the data without the correlations would result in an underestimate of the uncertainty of $a_{\mu}^{\rm HVP}$ and, potentially, a bias of its mean value. The construction of these covariance matrices must be statistically robust in order to ensure that they correctly describe the correlated relationship of the three measurements. 

The main purpose of this work is to formulate the covariance matrices required to determine the correct combination. In Section~\ref{Sec:KLOEmeasurements}, the three KLOE measurements of $\sigma\big(e^+e^-\rightarrow\pi^+\pi^-\gamma(\gamma)\big)$~\cite{Ambrosino:2008aa,KLOE08-KLOEnote,Ambrosino:2010bv,KLOE10-KLOEnote,Babusci:2012rp,KLOE12-KLOEnote} are reviewed and, in some cases, updated in order to ensure a consistent combination. Section~\ref{Sec:ConstructKLOECovMat} then focuses on the construction of the statistical and systematic covariance matrices for the combination of the KLOE measurements. In Section~\ref{Sec:KLOEcombination}, these matrices are then used to combine the three measurements into a single measurement of $\sigma^0\big(e^+e^-\rightarrow\pi^+\pi^-\gamma(\gamma)\big)$, which we use to provide an estimate of $a_{\mu}^{\rm \pi^+\pi^-}$. We then compare our results with the individual KLOE measurements and other experimental measurements of $\sigma_{\pi\pi(\gamma)}$.

\section{Measurements of $\sigma^0\big(e^+e^-\rightarrow\pi^+\pi^-\gamma(\gamma)\big)$ by the KLOE collaboration} \label{Sec:KLOEmeasurements}

\subsection{Determination of the $\pi^+\pi^-$ cross section}\label{Sec:pipixSec}

DA$\Phi$NE~\cite{DAFNE} is a high luminosity $e^+e^-$ collider that operates predominantly at the centre of mass energy equal to the $\phi$ meson mass, $\sqrt{s} = m_{\phi} = 1.0194 \text{ GeV} $~\cite{PDG2016}. The KLOE detector has been used to obtain measurements of the process $e^+e^-\rightarrow\pi^+\pi^-\gamma(\gamma)$ ~\cite{Ambrosino:2008aa,Ambrosino:2010bv,Babusci:2012rp,Venanzoni:2017ggn}. These measurements are achieved through radiative return, where the differential cross section is measured as a function of the invariant mass of the pion pair, $\sqrt{s'} = M_{\pi\pi}$. The cross section $\sigma_{\pi\pi} \equiv \sigma(e^+e^-\rightarrow\pi^+\pi^-)$ is then determined according to~\cite{Binner:1999bt} using the relation
\beq \label{pipidiffxSec}
s\frac{{\rm d}\sigma\big(\pi^+\pi^-\gamma\big)}{dM_{\pi\pi}^2} = \sigma_{\pi\pi}(M_{\pi\pi}^2)H(M_{\pi\pi}^2,s) \ ,
\eeq
where $H$ is the radiator function describing the emission of photons in the initial state~\cite{Rodrigo:2001kf,Czyz:2002np,Czyz:2003ue,Czyz:2004rj}. Equation~\eqref{pipidiffxSec} is valid neglecting the contribution from FSR, although it is properly accounted for in the KLOE analyses~\cite{Ambrosino:2008aa,Ambrosino:2010bv,Babusci:2012rp,Actis:2010gg}.

The KLOE collaboration have performed three measurements of the cross section $\sigma\big(e^+e^-\allowbreak\rightarrow\pi^+\pi^-\gamma(\gamma)\big)$~\cite{Ambrosino:2008aa,KLOE08-KLOEnote,Ambrosino:2010bv,KLOE10-KLOEnote,Babusci:2012rp,KLOE12-KLOEnote}. All three published cross sections are bare (undressed of all VP effects) and including FSR. For the first two, which for the purposes of this study we shall denote as KLOE08~\cite{Ambrosino:2008aa} and KLOE10~\cite{Ambrosino:2010bv}, the bare cross section is obtained by~\cite{FJ16VP} 
\beq \label{barexSec}
\sigma^{0}_{\pi\pi(\gamma)}(s') = \sigma_{\pi\pi(\gamma)}(s') \abs{1- \Pi(s')}^2 ,
\eeq
where the superscript `$0$' indicates that the cross section is bare, the subscript $(\gamma)$ indicates that the cross section includes FSR, $ \sigma_{\pi\pi(\gamma)}(s')$ is obtained using equation~\eqref{pipidiffxSec} and $\Pi(s')$ is the vacuum polarisation containing both real and imaginary parts~\cite{KLOE-2:2016mgi}.\footnote{The correction used previously for KLOE08 ~\cite{Ambrosino:2008aa,KLOE08-KLOEnote} and KLOE10~\cite{Ambrosino:2010bv,KLOE10-KLOEnote} contained only the real part of the VP. This has been updated in this analysis to incorporate the full VP with both the real and imaginary parts, where the imaginary part is small and sub-leading compared to the real contribution (see Section~\ref{KLOEmeasurements} for more details).}

For the third measurement of $\sigma^{0}_{\pi\pi(\gamma)}(s')$, namely KLOE12~\cite{Babusci:2012rp}, a reciprocal relation to equation~\eqref{pipidiffxSec} was utilised, allowing for a bin-by-bin normalisation of the $\pi^+\pi^-$ cross section by the $\mu^+\mu^-$ cross section. For the same invariant mass squared, the ratio of the $\pi^+\pi^-\gamma$ and $\mu^+\mu^-\gamma$ differential cross sections allows the relation
\beq \label{KLOE12xSec}
\sigma^0_{\pi\pi(\gamma)}(s') = \frac{{\rm d}\sigma\big(\pi^+\pi^-\gamma\big)/{\rm d}s'}{{\rm d}\sigma\big(\mu^+\mu^-\gamma)\big/{\rm d}s'} \times \sigma^0_{(\gamma)}(e^+e^-\rightarrow\mu^+\mu^-,s') \ ,
\eeq
where $s' = M_{\pi\pi}^2 = M_{\mu\mu}^2$. This normalisation has many advantages concerning the determination of the cross section. Importantly, the ratio in equation~\eqref{KLOE12xSec} benefits from the cancellation of the radiator function for initial state radiation (ISR) and of the VP correction, manifestly resulting in a bare cross section. Therefore, the undressing procedure described by equation~\eqref{barexSec} is not applied to KLOE12, although the FSR contribution to the $\pi^+\pi^-$ production must again be included. 

The pion form factor, $\left|F_{\pi}\right|^2$, is determined for all three measurements to be
\beq
\left|F_{\pi}(s')\right|^2 = \frac{3}{\pi}\frac{s'}{\alpha^2\beta_{\pi}^3(s')}\frac{\sigma^0_{\pi\pi(\gamma)}(s')}{\abs{1- \Pi(s')}^2}\Big(1-\frac{\alpha}{\pi}\eta_{\pi}(s')\Big) \ ,
\eeq
where $\alpha \equiv \alpha(0)$, $\beta_{\pi}(s') = \sqrt{1-4m_{\pi}^2/s'}$ and $\eta_{\pi}$ is the inclusive FSR correction assuming point-like pions~\cite{Hoefer:2001mx}.
 
\subsection{The KLOE measurements} \label{KLOEmeasurements}

The experimental analysis of each KLOE measurement of $\sigma\big(e^+e^-\rightarrow\pi^+\pi^-\gamma(\gamma)\big)$ has been reviewed and, in some cases, updated in order to ensure a more precise and consistent combination of the three measurements. In the following, each measurement is discussed individually, where any changes to the respective analysis are explicitly stated. 

The KLOE08 measurement consists of 60 data points in the range $0.35< s' <0.95$ GeV$^2$, covering the dominant $\rho$ resonance structure and the $\rho-\omega$ interference region in the $\pi^+\pi^-$ final state. The uncertainties of the cross section are dominated by the systematics uncertainties, especially in the region where the cross section is large. The KLOE08 data have been updated with respect to~\cite{Ambrosino:2008aa} to incorporate the following necessary changes:
\begin{itemize}
\item The data have been undressed of VP effects using an updated routine~\cite{FJ16VP} compared to the one used previously~\cite{FJ03VP}, which now corrects the data using a more appropriate energy grid parametrisation for the determination of the VP.
\item The VP correction contains both real and imaginary parts, whereas previously the data were only corrected for the real part of the VP.
\item The data are not rounded as they were in~\cite{Ambrosino:2008aa} to ensure that the statistical and systematic uncertainties correspond to the variances that enter into the diagonal elements of the corresponding covariance matrices.
\item The calculation of the cross section has been updated with respect to the precision of input parameters and fundamental constants~\cite{PDG2016}.
\end{itemize}
We find, using the updated data for KLOE08, a contribution to the anomalous magnetic moment of the muon of
\beq
a_{\mu}^{\pi^+\pi^-}({\rm KLOE08}, 0.35< s' <0.95 \text{ GeV}^2) = (386.6 \pm 0.4_{\rm stat} \pm 3.3_{\rm sys}) \times 10^{-10} ,
\eeq
which exhibits a decrease in the mean value of $a_{\mu}^{\pi^+\pi^-}$ when compared to the estimate quoted in~\cite{Ambrosino:2008aa} that is largely due to the updated determination of the VP.
The updated cross section and pion form factor vectors with corresponding covariance matrices for the statistical and systematic uncertainties are available from~\cite{KLOE17ppg}.

The KLOE10 measurement totals 75 data points in the range $0.1< s' <0.85$ GeV$^2$. This analysis~\cite{Ambrosino:2010bv} selected events that included a photon detected in the calorimeter at large polar angle, allowing the measurement to be taken at lower $s'$ closer to threshold. The fifty energy bins of the data in the range $0.35< s' < 0.85$ GeV$^2$ are identical to the fifty KLOE08 bins in the same interval. The KLOE10 cross section has been updated in the same way as KLOE08, with the application of the improved VP correction~\cite{FJ16VP}, the non-rounded data and improved parameter precision resulting in 
\beq
a_{\mu}^{\pi^+\pi^-}({\rm KLOE10}, 0.10 < s' <0.85 \text{ GeV}^2) = (477.9 \pm 2.0_{\rm stat} \pm 6.7_{\rm sys}) \times 10^{-10} ,
\eeq
which, like observed with KLOE08, results in a decrease in the mean value of $a_{\mu}^{\pi^+\pi^-}$ compared to the estimate in~\cite{Ambrosino:2010bv}. The updated KLOE10 data vectors and covariance matrices are available from~\cite{KLOE17ppg}.

\begin{figure}[!t] 
  \begin{center}  
  \begin{tikzpicture}[>=Latex,every node/.style={scale=1.5,outer sep=0,font={\tiny \bf \sffamily}}]
       
        \begin{scope}[x={(14,0)},y={(0,13.25)}]
           
                
         \node(K08_1)[draw, black, thick, rounded corners=2.5pt, align=center, inner sep=2.5pt] at (0.165,0.955) {Observed Spectrum for\\ $\pi\pi\gamma(\gamma)$ events};

         \node(K08_2)[draw, black, thick, align=center, inner sep=1pt] at (0.165,0.885) {(Level 3 Trigger)};
         
         \node(K08_3)[draw, black, thick, align=center, inner sep=1pt] at (0.165,0.84) {Offline filter corr.};

          \node(K08_4)[draw, black, thick, align=center, inner sep=1.5pt] at (0.165,0.79) {Background subtr.};       

          \node(K08_5)[draw, black, thick, align=center, inner sep=1.5pt] at (0.165,0.745) {${\bf \mathsf{M_{Trk}}}$+${\bf \Delta\mathsf{E_{Miss}}}$ corr.};

          \node(K08_6)[draw, black, thick, align=center, inner sep=1.5pt] at (0.165,0.69) {Unfolding (${\bf \mathsf{M^2_{Rec}}\to \mathsf{M^2_{True}}}$)};

          \node(K08_7)[draw, black, thick, align=center, inner sep=1.5pt] at (0.165,0.64) {Corr. for border eff. in Acc.};

          \node(K08_8)[draw, black, thick, align=center, inner sep=1.5pt] at (0.165,0.59) {$\pi/\mathsf{e}$ likelihood + TCA corr.};

          \node(K08_9)[draw, black, thick, align=center, inner sep=1.5pt] at (0.165,0.535) {Tracking corr.};

          \node(K08_10)[draw, black, thick, align=center, inner sep=1.5pt] at (0.165,0.485) {Trigger corr.};          

          \node(K08_11)[draw, black, thick, align=center, inner sep=1.5pt] at (0.165,0.435) {Unshifting (${\bf \mathsf{M^2_{\pi\pi}}\to \mathsf{M^2_{\gamma^*}}}$)};

          \node(K08_12)[draw, black, thick, align=center, inner sep=1.5pt] at (0.165,0.38) {Acceptance $\theta_\pi$ corr.};  

          \node(K08_13)[draw, black, thick, align=center, inner sep=1.5pt] at (0.165,0.27) {Luminosity corr.};

          \node(K08_14)[draw, black, thick, align=center, inner sep=1.5pt] at (0.165,0.21) {Acceptance $\theta_\Sigma$ corr.};

          \node(K08_15)[draw, black, thick, align=center, inner sep=1.5pt] at (0.165,0.16) {Division by Radiator H};

          \node(K08_16)[draw, black, thick, align=center, inner sep=1.5pt] at (0.165,0.1) {Corr. for vac. pol.};

         \node(K08_17)[draw, black, thick, rounded corners=1.5pt, align=center, inner sep=2.5pt] at (0.165,0.04) {$\sigma_{\pi\pi}$ KLOE08};          
          
         \draw[-{>[scale=.5]},black,thick] (K08_1) -- (K08_2);
         \draw[-{>[scale=.5]},black,thick] (K08_2) -- (K08_3);
         \draw[-{>[scale=.5]},black,thick] (K08_3) -- (K08_4);
         \draw[-{>[scale=.5]},black,thick] (K08_4) -- (K08_5);
         \draw[-{>[scale=.5]},black,thick] (K08_5) -- (K08_6);
         \draw[-{>[scale=.5]},black,thick] (K08_6) -- (K08_7);
         \draw[-{>[scale=.5]},black,thick] (K08_7) -- (K08_8);
         \draw[-{>[scale=.5]},black,thick] (K08_8) -- (K08_9);
         \draw[-{>[scale=.5]},black,thick] (K08_9) -- (K08_10);
         \draw[-{>[scale=.5]},black,thick] (K08_10) -- (K08_11);
         \draw[-{>[scale=.5]},black,thick] (K08_11) -- (K08_12);
         \draw[-{>[scale=.5]},black,line width = 3pt] (K08_12) -- (K08_13);
         \draw[-{>[scale=.5]},black,thick] (K08_13) -- (K08_14);
         \draw[-{>[scale=.5]},black,thick] (K08_14) -- (K08_15);
         \draw[-{>[scale=.5]},black,thick] (K08_15) -- (K08_16);
         \draw[-{>[scale=.5]},black,thick] (K08_16) -- (K08_17);

        
         \node(K12_1)[draw, black, thick, rounded corners=2.5pt, align=center, inner sep=2.5pt] at (0.5,0.955) {Observed Spectrum for\\ $\mu\mu\gamma(\gamma)$ events};

         \node(K12_2)[draw, black, thick, align=center, inner sep=1pt] at (0.5,0.885) {(Level 3 Trigger)};
         
         \node(K12_3)[draw, black, thick, align=center, inner sep=1pt] at (0.5,0.84) {Offline filter corr.};

          \node(K12_4)[draw, black, thick, align=center, inner sep=1.5pt] at (0.5,0.79) {Background subtr.};       

          \node(K12_5)[draw, black, thick, align=center, inner sep=1.5pt] at (0.5,0.745) {Corr. for border eff. in Acc.};
          
          \node(K12_6)[draw, black, thick, align=center, inner sep=1.5pt] at (0.5,0.69) {${\bf \mathsf{M_{Trk}}}$+${\bf \Delta\mathsf{E_{Miss}}}$ corr.};

          \node(K12_7)[draw, black, thick, align=center, inner sep=1.5pt] at (0.5,0.64) {Unfolding (${\bf \mathsf{M^2_{Rec}}\to \mathsf{M^2_{True}}}$)};

          \node(K12_8)[draw, black, thick, align=center, inner sep=1.5pt] at (0.5,0.59) {$\pi/\mathsf{e}$ likelihood + TCA corr.};

          \node(K12_9)[draw, black, thick, align=center, inner sep=1.5pt] at (0.5,0.535) {Tracking corr.};

          \node(K12_10)[draw, black, thick, align=center, inner sep=1.5pt] at (0.5,0.485) {Trigger corr.};          

          \node(K12_11)[draw, black, thick, align=center, inner sep=1.5pt] at (0.5,0.435) {Acceptance $\theta_\pi$ corr.};  

          \node(K12_12)[draw, black, thick, align=center, inner sep=1.5pt] at (0.5,0.38) {FSR $\to$ ISR corr.};

          \node(K12_13)[draw, black, thick, align=center, inner sep=1.5pt] at (0.5,0.21) {Acceptance corr.\\ ($\theta_\Sigma^{\mu\mu,\bf{\mathsf{ISR}}}/\theta_\Sigma^{\pi\pi,\bf{\mathsf{ISR}}}$)};

         \node(K12_14)[draw, black, thick, rounded corners=1.5pt, align=center, inner sep=2.5pt] at (0.5,0.04) {$\sigma_{\pi\pi}$ KLOE12};          
          
         \draw[-{>[scale=.5]},black,thick] (K12_1) -- (K12_2);
         \draw[-{>[scale=.5]},black,thick] (K12_2) -- (K12_3);
         \draw[-{>[scale=.5]},black,thick] (K12_3) -- (K12_4);
         \draw[-{>[scale=.5]},black,thick] (K12_4) -- (K12_5);
         \draw[-{>[scale=.5]},black,thick] (K12_5) -- (K12_6);
         \draw[-{>[scale=.5]},black,thick] (K12_6) -- (K12_7);
         \draw[-{>[scale=.5]},black,thick] (K12_7) -- (K12_8);
         \draw[-{>[scale=.5]},black,thick] (K12_8) -- (K12_9);
         \draw[-{>[scale=.5]},black,thick] (K12_9) -- (K12_10);
         \draw[-{>[scale=.5]},black,thick] (K12_10) -- (K12_11);
         \draw[-{>[scale=.5]},black,thick] (K12_11) -- (K12_12);         
         \draw[black,thick] (K12_12) -- (K12_13);
         \draw[-{>[scale=.5]},black,thick] (K12_13) -- (K12_14);

         \draw[black,line width = 3pt] (0.165,0.335) -- (0.5,0.335);
         \draw[-{>[scale=.5]},black,line width = 3pt] (0.5,0.3395) -- (K12_13);
         \node(K0812)[anchor=west,align=left] at (0.51,0.32) {Division\\$\pi\pi\gamma(\gamma)/\mu\mu\gamma(\gamma)$};
     
        
         \node(K10_1)[draw, black, thick, rounded corners=2.5pt, align=center, inner sep=2.5pt] at (0.85,0.955) {Observed Spectrum for\\ $\pi\pi\gamma(\gamma)$ events};

         \node(K10_2)[draw, black, thick, align=center, inner sep=1pt] at (0.85,0.885) {(Level 3 Trigger)};
         
         \node(K10_3)[draw, black, thick, align=center, inner sep=1pt] at (0.85,0.84) {Offline filter corr.};

          \node(K10_4)[draw, black, thick, align=center, inner sep=1.5pt] at (0.85,0.79) {Background subtr.};       

          \node(K10_5)[draw, black, thick, align=center, inner sep=1.5pt] at (0.85,0.735) {Trigger corr.};

          \node(K10_6)[draw, black, thick, align=center, inner sep=1.5pt] at (0.85,0.69) {$\pi/\mathsf{e}$ likelihood + TCA corr.};
        
          \node(K10_7)[draw, black, thick, align=center, inner sep=1.5pt] at (0.85,0.64) {Unfolding (${\bf \mathsf{M^2_{Rec}}\to \mathsf{M^2_{True}}}$)};

          \node(K10_8)[draw, black, thick, align=center, inner sep=1.5pt] at (0.85,0.59) {f${}_0$+$\rho\pi$ corr.};

          \node(K10_9)[draw, black, thick, align=center, inner sep=1.5pt] at (0.85,0.535) {Tracking corr. data/MC};

          \node(K10_10)[draw, black, thick, align=center, inner sep=1.5pt] at (0.85,0.485) {Photon corr. data/MC};      

          \node(K10_11)[draw, black, thick, align=center, inner sep=1.5pt] at (0.85,0.435) {Global MC efficiency};           
          
          \node(K10_12)[draw, black, thick, align=center, inner sep=1.5pt] at (0.85,0.38) {Unshifting (${\bf \mathsf{M^2_{\pi\pi}}\to \mathsf{M^2_{\gamma^*}}}$)};

          \node(K10_13)[draw, black, thick, align=center, inner sep=1.5pt] at (0.85,0.27) {Luminosity corr.};

          \node(K10_14)[draw, black, thick, align=center, inner sep=1.5pt] at (0.85,0.21) {Division by Radiator H};

          \node(K10_15)[draw, black, thick, align=center, inner sep=1.5pt] at (0.85,0.16) {Corr. for vac. pol.};

         \node(K10_16)[draw, black, thick, rounded corners=1.5pt, align=center, inner sep=2.5pt] at (0.85,0.04) {$\sigma_{\pi\pi}$ KLOE10};          
          
         \draw[-{>[scale=.5]},black,thick] (K10_1) -- (K10_2);
         \draw[-{>[scale=.5]},black,thick] (K10_2) -- (K10_3);
         \draw[-{>[scale=.5]},black,thick] (K10_3) -- (K10_4);
         \draw[-{>[scale=.5]},black,thick] (K10_4) -- (K10_5);
         \draw[-{>[scale=.5]},black,thick] (K10_5) -- (K10_6);
         \draw[-{>[scale=.5]},black,thick] (K10_6) -- (K10_7);
         \draw[-{>[scale=.5]},black,thick] (K10_7) -- (K10_8);
         \draw[-{>[scale=.5]},black,thick] (K10_8) -- (K10_9);
         \draw[-{>[scale=.5]},black,thick] (K10_9) -- (K10_10);
         \draw[-{>[scale=.5]},black,thick] (K10_10) -- (K10_11);
         \draw[-{>[scale=.5]},black,thick] (K10_11) -- (K10_12);
         \draw[-{>[scale=.5]},black,thick] (K10_12) -- (K10_13);
         \draw[-{>[scale=.5]},black,thick] (K10_13) -- (K10_14);
         \draw[-{>[scale=.5]},black,thick] (K10_14) -- (K10_15);
         \draw[-{>[scale=.5]},black,thick] (K10_15) -- (K10_16);
         
         \end{scope}
         
  \end{tikzpicture}
\caption{The flow of the experimental analyses of all three $\sigma^0\big(e^+e^-\rightarrow\pi^+\pi^-\gamma(\gamma)\big)$ cross section measurements. The point where the KLOE08 $\pi^+\pi^-\gamma(\gamma)$ data enter the KLOE12 analysis is indicated by the bold black arrows.}    \label{KLOEanalysisflow}
\end{center}
\end{figure}
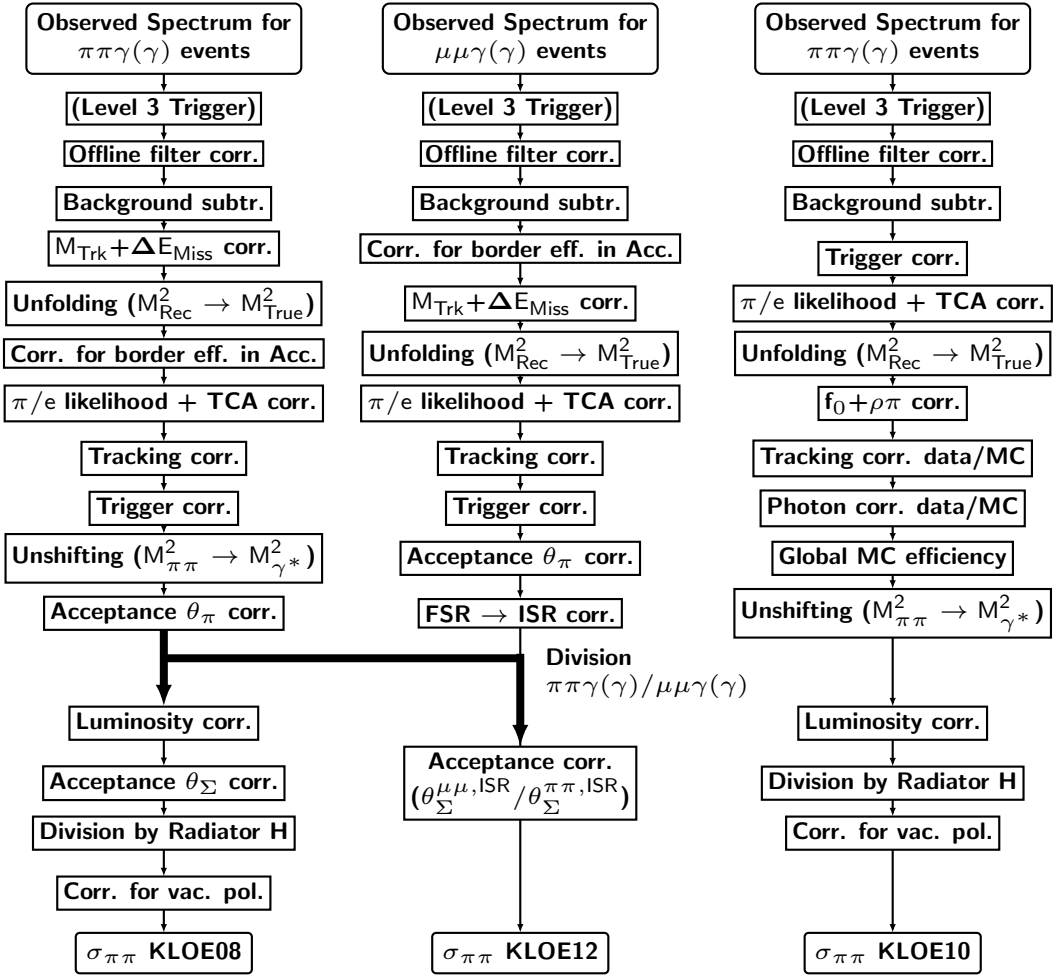

The KLOE12 measurement was determined as a $\mu^+\mu^-\gamma$ normalised cross section, as described briefly in Section~\ref{Sec:pipixSec}. The $\mu^+\mu^-$ cross section was measured for the analysis, whereas the KLOE08 $\pi^+\pi^-$ data were used as the input into equation~\eqref{KLOE12xSec}, with the KLOE12 measurement having an identical binning and energy range to KLOE08. As these measurements share the same two-pion data, KLOE08 and KLOE12 are highly correlated and it is imperative that they be treated as such in any combination of the two measurements. The KLOE12 cross section has been updated with respect to the use of non-rounded data and input parameter precision. The ratio in equation~\eqref{KLOE12xSec} benefits from the cancellation of the VP correction and, therefore, does not require an updated VP correction as with the KLOE08 and KLOE10 cross section data. For the contribution to the muon magnetic anomaly, from the KLOE12 data alone, we find
\beq \label{KLOE12}
a_{\mu}^{\pi^+\pi^-}({\rm KLOE12}, 0.35< s' <0.95 \text{ GeV}^2) = (385.1 \pm 1.2_{\rm stat} \pm 2.3_{\rm sys}) \times 10^{-10} .
\eeq
Here, the error has reduced since~\cite{Babusci:2012rp}, where a flaw in the previous error calculation resulted in an overestimation of the published systematic uncertainty and, as a result, there have been necessary changes to the KLOE12 systematic matrix construction.\footnote{The KLOE12 systematic uncertainty has reduced from $2.7\times10^{-10}$ given in~\cite{Babusci:2012rp} to $2.3\times10^{-10}$ in this analysis.} The updated KLOE12 data are available from~\cite{KLOE17ppg}.

\section{Constructing the KLOE combination covariance matrices} \label{Sec:ConstructKLOECovMat}

The flow of the experimental analyses for the KLOE08, KLOE10 and KLOE12 measurements is shown in Figure~\ref{KLOEanalysisflow}. In the case of the KLOE12 measurement, the beginning of the flow refers to the measurement of $\mu^+\mu^-\gamma(\gamma)$. The point where the KLOE08 $\pi^+\pi^-\gamma(\gamma)$ data enters is clearly marked. This diagram exhibits the extent of the correlation between KLOE08 and KLOE12, with correlations existing for all elements of the KLOE08 $\sigma_{\pi\pi}$ analysis from the observed spectrum of $\pi^+\pi^-\gamma(\gamma)$ events up to the acceptance correction. In addition, the degree of correlation between KLOE08 and KLOE10 or KLOE10 and KLOE12 is clearly shown, with many parts of the experimental analyses being common to a pair of measurements or having been obtained through a similar method. 

\begin{figure}[t] 
\beq
\small
\left(\begin{array}{ccc|ccc|ccc}
\cdots&\cdots &\cdots&\cdots&\cdots &\cdots&\text{ }\cdots&\cdots&\cdots \\
\cdots&\cdots &\cdots&\cdots&\cdots &\cdots&\text{ }\cdots&\cdots&\cdots \\
\multicolumn{3}{c}{\text{KLOE08}} \vline & \cdots&\multicolumn{1}{c}{\text{KLOE0810}}& \cdots &  \multicolumn{3}{c}{\text{KLOE0812}} \\
\multicolumn{3}{c}{60\times60} \vline &\cdots&60\times75&\cdots &\multicolumn{3}{c}{60\times60} \\
\cdots&\cdots &\cdots&\cdots&\cdots &\cdots&\text{ }\cdots&\cdots&\cdots \\
\cdots&\cdots &\cdots&\cdots&\cdots &\cdots&\text{ }\cdots&\cdots&\cdots \\
\hline
\cdots&\cdots &\cdots&\cdots&\cdots &\cdots&\text{ }\cdots&\cdots&\cdots \\
\cdots&\cdots &\cdots&\cdots&\cdots &\cdots&\text{ }\cdots&\cdots&\cdots \\
\cdots&\cdots &\cdots&\cdots&\cdots &\cdots&\text{ }\cdots&\cdots&\cdots \\
\multicolumn{3}{c}{\text{KLOE1008}} \vline & \cdots&\multicolumn{1}{c}{\text{KLOE10}}& \cdots &  \multicolumn{3}{c}{\text{KLOE1012}} \\
\multicolumn{3}{c}{75\times60} \vline &\cdots&75\times75&\cdots &\multicolumn{3}{c}{75\times60} \\
\cdots&\cdots &\cdots&\cdots&\cdots &\cdots&\text{ }\cdots&\cdots&\cdots \\
\cdots&\cdots &\cdots&\cdots&\cdots &\cdots&\text{ }\cdots &\cdots&\cdots\\
\cdots&\cdots &\cdots&\cdots&\cdots &\cdots&\text{ }\cdots &\cdots&\cdots\\
\hline
\cdots&\cdots &\cdots&\cdots&\cdots &\cdots&\text{ }\cdots&\cdots&\cdots \\
\cdots&\cdots &\cdots&\cdots&\cdots &\cdots&\text{ }\cdots&\cdots&\cdots \\
\multicolumn{3}{c}{\text{KLOE1208}} \vline &\cdots&\multicolumn{1}{c}{\text{KLOE1210}}&\cdots& \multicolumn{3}{c}{\text{KLOE12}} \\
\multicolumn{3}{c}{60\times60} \vline &\cdots&60\times75&\cdots &\multicolumn{3}{c}{60\times60} \\\
\cdots&\cdots &\cdots&\cdots&\cdots &\cdots&\text{ }\cdots&\cdots&\cdots \\
\cdots&\cdots &\cdots&\cdots&\cdots &\cdots&\text{ }\cdots&\cdots&\cdots \\
\end{array}\right) \ \nonumber
\eeq
     \caption{\small The KLOE $\pi^+\pi^-\gamma(\gamma)$ combination matrix structure for both the statistical and systematic covariance matrices.}     \label{KLOEmatrix}
\end{figure} 
 \begin{figure}[htbp]
\centering
\vspace{-1cm}
\hspace{-0.5cm}
 {\subfloat[Statistical correlation matrix]{%
    \includegraphics[width= 0.8\textwidth, trim={0 0 0 1.15cm}, clip]{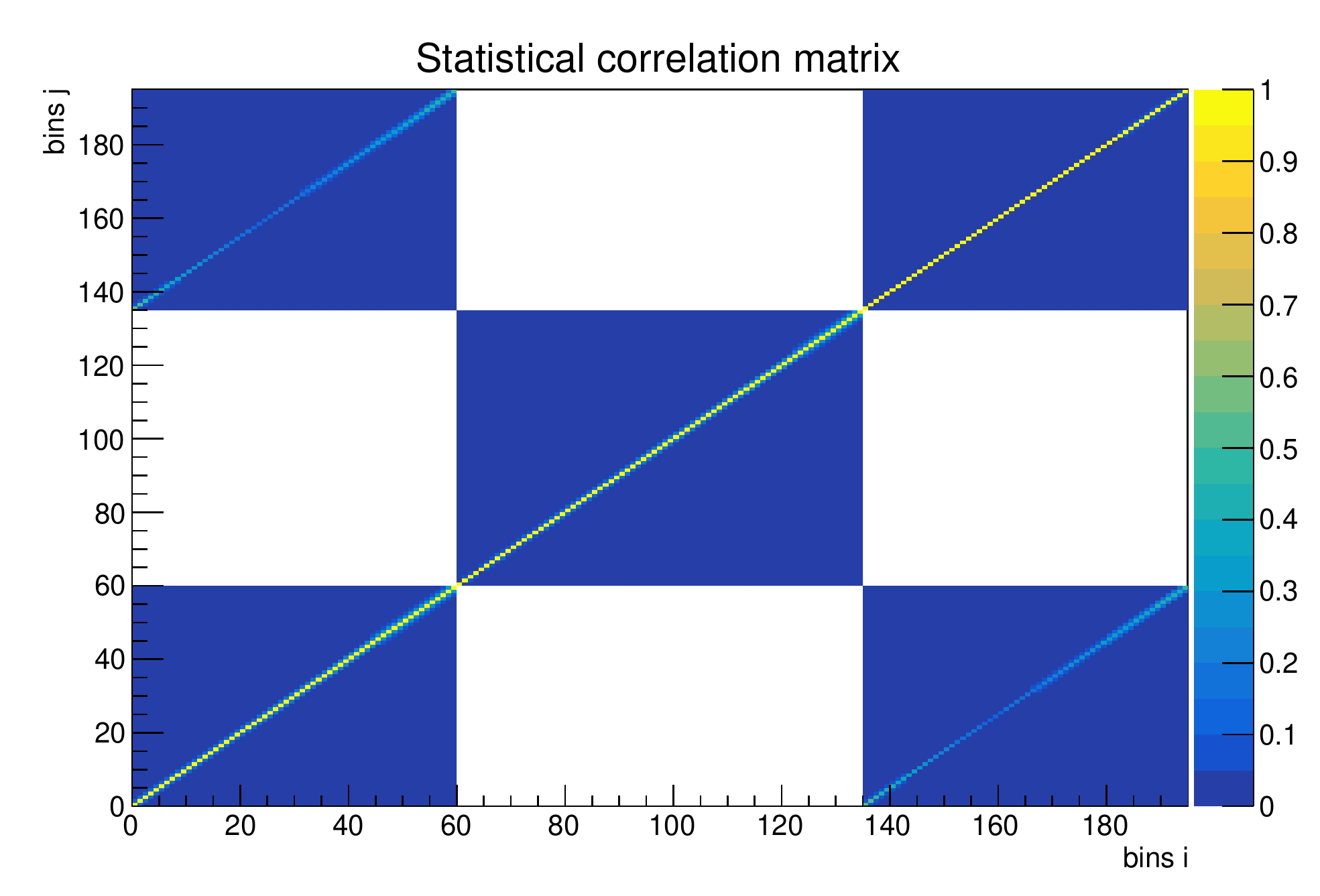}}    \hspace{0.4cm} 
  \subfloat[Systematic correlation matrix]{%
    \includegraphics[width= 0.8\textwidth, trim={0 0 0 1.15cm}, clip]{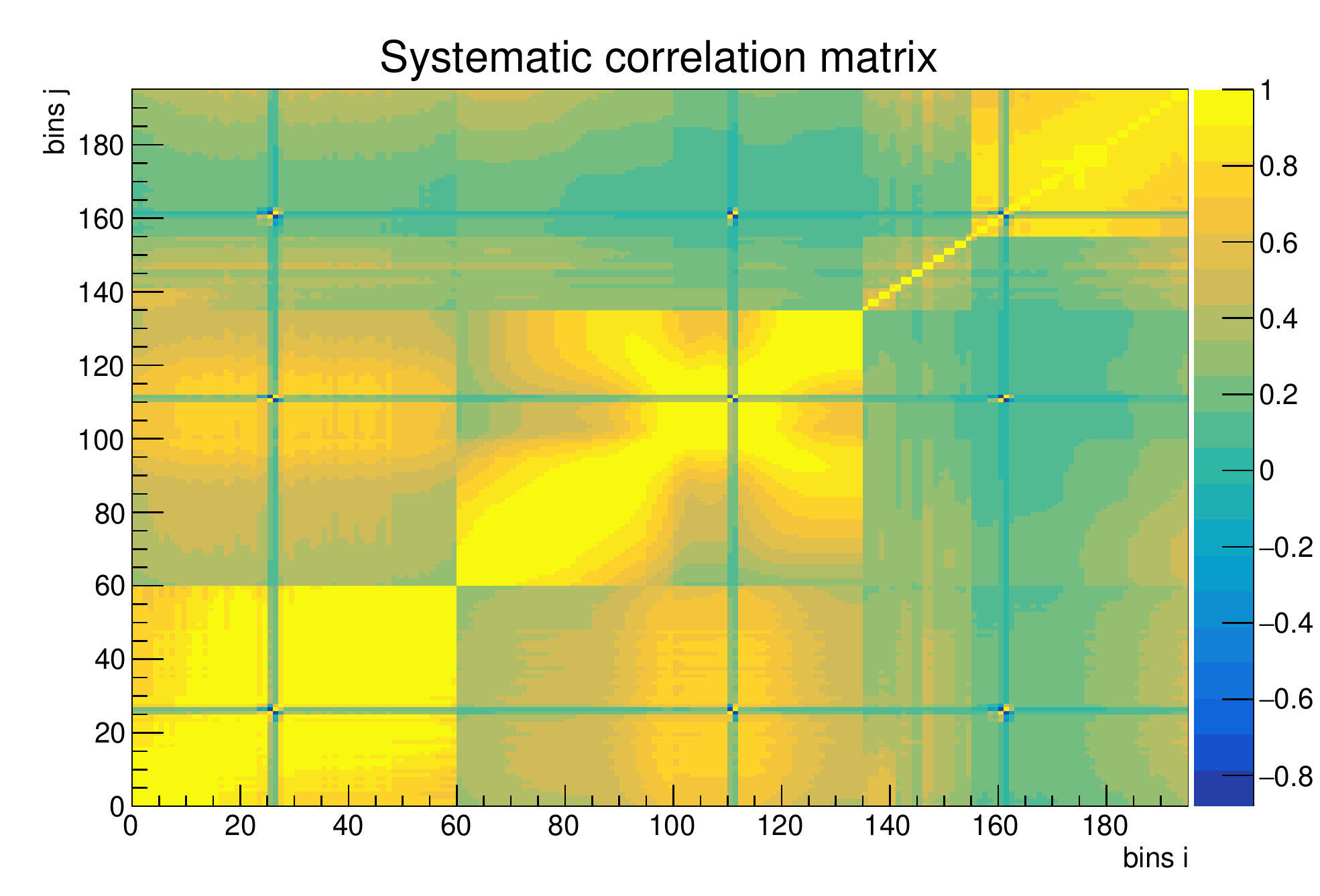}}\hfill 
  \caption{\small The correlation structure of the 195x195 statistical and systematic combination matrices. In each case, the axis on the right represents the overall correlation coefficient ($\rho_{ij} =-1 \leq \rho \leq 1$), where the corresponding colour indicates the degree of correlation at each point in the respective matrix (colour online).}\label{KLOEcorrelation} }
\end{figure} 

The KLOE statistical and systematic combination covariance matrices are $195\times195$ matrices and are depicted in Figure~\ref{KLOEmatrix}.  They have been carefully constructed to satisfy the necessary mathematical properties of a covariance matrix, details of which are described in Appendix~\ref{App:CovMat}. The KLOE08, KLOE10 and KLOE12 diagonal blocks are simply the covariance matrices of the individual measurements. The KLOE0810 block describes the correlation between KLOE08 and KLOE10, with corresponding definitions for KLOE0812 and KLOE1012. Statistical uncertainties are, in general, uncorrelated ($\rho^{\alpha_{\rm stat}}_{ij}|_{i\neq j} = 0$, where $\rho_{ij}$ is the correlation coefficient defined in Appendix~\ref{App:CovMat}) and only contribute to the diagonal elements of the corresponding correlation block of the combination covariance matrix. The exceptions to this are the unfolding and unshifting corrections\footnote{While the unfolding correction accounts for the smearing due to the detector resolution, the unshifting is a redistribution correction that accounts for photons emitted through final state radiation, which results in shifting the observed value of $s'$ away from the squared invariant mass of the virtual photon $s_\gamma^*$~\cite{Actis:2010gg}.}  (see Section~\ref{stat}), which both contribute to the non-diagonal elements of the statistical matrix ($\rho^{\alpha_{\rm stat}}_{ij} =-1 \leq \rho \leq 1$). For systematic (sys) uncertainties, all data points are taken to be 100\% correlated or anti-correlated ($\rho^{\alpha_{\rm sys}}_{ij} = \pm1$). The resulting correlation structures for both the statistical and systematic matrix are shown in Figure~\ref{KLOEcorrelation}. In the following, we outline the correlations that exist for and between the individual measurements for the statistical and systematic uncertainties separately. 

\subsection{Statistical correlations} \label{stat}

Other than those that exist as part of the individual analyses for the KLOE08, KLOE10 and KLOE12 diagonal sub-matrices in the statistical $195\times195$ combination covariance matrix depicted in Figure~\ref{KLOEmatrix}, the only statistical correlations that are present are those due to the two-pion data that are shared between KLOE08 and KLOE12. These occupy the KLOE0812 and KLOE1208 blocks of the statistical combination covariance matrix. As no statistical correlations exist between KLOE08 and KLOE10 or KLOE10 and KLOE12, all elements of the KLOE0810 (KLOE1008) and KLOE1012 (KLOE1210) correlation blocks of the statistical covariance matrix are zero. This can be seen diagrammatically in Figure~\ref{KLOEcorrelation}.

The individual KLOE08, KLOE10 and KLOE12 statistical covariance matrices (corresponding to the diagonal blocks of the statistical combination matrix given by Figure~\ref{KLOEmatrix}) describe all statistical uncertainties inherent in the respective experimental analysis. The contributions to the statistical covariance matrices from the unfolding and unshifting procedures are partially correlated, where the correlation coefficients are defined by the unfolding~\cite{SM13RMC,DAgostini:1994fjx,D'AgostiniUndfolding2} and unshifting~\cite{SM13RMC} procedures themselves. Details regarding these procedures and all other statistical uncertainties (which are considered to be fully uncorrelated) can be found in~\cite{KLOE08-KLOEnote,KLOE10-KLOEnote,KLOE12-KLOEnote}.

The KLOE0812 statistical correlation block receives contributions from all corrections to the KLOE08 $\pi^+\pi^-\gamma(\gamma)$ data up until the point where these data enter the KLOE12 analysis. Following the experimental analysis flow for KLOE08 in Figure~\ref{KLOEanalysisflow}, these include the detector resolution correction (unfolding), the correction for border efficiency in the acceptance,  the pion identification efficiency ($\pi/\varepsilon$ likelihood), the tracking efficiency, the trigger corrections, the unshifting of $M_{\pi\pi}^2 \rightarrow (M_{\pi\pi}^0)^2$ and the acceptance for the cuts in $\theta_{\pi}$ and $\theta_{\pi\pi}$~\cite{KLOE08-KLOEnote}. All corrections prior to the unfolding in the analysis flow are included in the unfolded KLOE08 $\pi^+\pi^-\gamma(\gamma)$ spectrum and, therefore, manifestly enter the KLOE0812 correlations through the correlations of the unfolding.  As the unfolding (unf) and unshifting (uns) corrections are identically correlated for the KLOE08 and KLOE12 statistical covariance matrices, these correlations must be reflected in the KLOE0812 correlation block exactly in the form
\beq
\rho^{0812,{\rm unf/uns}}_{ij} = \rho^{1208,{\rm unf/uns}}_{ji} = \rho^{08,{\rm unf/uns}}_{ij} = \rho^{12,{\rm unf/uns}}_{ij}.
\eeq
 Not doing so would result in the statistical covariance matrix having negative eigenvalues, therefore violating the condition that the covariance matrix is a positive semi-definite matrix. All remaining correlated statistical uncertainties only enter into the diagonal elements of the KLOE0812 correlation block, as they are fully correlated only for the same energy bins between the two measurements.

\subsection{Systematic correlations} \label{sys}

All correlation blocks in Figure~\ref{KLOEmatrix} receive contributions from systematic uncertainties, as can be seen clearly in Figure~\ref{KLOEcorrelation}. Unless stated otherwise, for any two bins $i$ and $j$, systematic uncertainties where correlations exist are fully correlated ($\rho_{ij} = +1$) or anti-correlated ($\rho_{ij} = -1$).

For the individual measurements, apart from two exceptions, all sources of systematic uncertainty are fully correlated between all energy bins. The first exception is the systematic uncertainty due to the unfolding, which only contributes at the sharp descent of the cross section in the $\rho-\omega$ interference region.  Here, an identical unfolding uncertainty enters for five bins of the KLOE08 and KLOE12 analyses and is anti-correlated only for pairs of bins that are on different sides of this sharp descent of the cross section.  For KLOE10, the only two affected bins are those directly before and directly after the sharp descent in the cross section, where the uncertainties are fully anti-correlated between these two bins. The second exception is the weighted background subtraction for KLOE12, where in the experimental analysis the weights of the fitted $e^+e^-\gamma$, $\pi\pi\gamma$ and $\pi\pi\pi$ backgrounds to the $\mu^+\mu^-\gamma(\gamma)$ spectrum are distributed over neighbouring two-bin intervals from 0.32 to 0.96 GeV$^2$. For the KLOE12 systematic covariance matrix, this results in only neighbouring bins from 0.36 to 0.94 GeV$^2$ being correlated with each other for this background subtraction uncertainty, where the first and last bin remain entirely uncorrelated in this case. A comprehensive discussion concerning this and all other systematic uncertainties for each measurement can be found in~\cite{KLOE08-KLOEnote,KLOE10-KLOEnote,KLOE12-KLOEnote}.

Importantly, for the KLOE12 systematic covariance matrix the trigger, L3 (software trigger), trackmass, tracking efficiency, acceptance and background subtraction corrections are applied to both the $\pi^+\pi^-\gamma$ and $\mu^+\mu^-\gamma$ data that enter into the ratio in equation~\eqref{KLOE12xSec} and, therefore, the corresponding uncertainties from a given source between the $\pi^+\pi^-\gamma$ and $\mu^+\mu^-\gamma$ data are correlated.\footnote{This only refers to the correlation of uncertainties from a specific source between the $\pi^+\pi^-\gamma$ analysis and the $\mu^+\mu^-\gamma$ analysis that enter into the KLOE12 ratio. The correlation between the KLOE08 $\pi^+\pi^-\gamma$ data and the KLOE12 cross section ratio are described in detail in the discussion of the KLOE0812 block of the systematic covariance matrix.} Formally, the ratio of these correction uncertainties results in a reduction of the total uncertainty of $a_\mu^{\pi^+\pi^-}$, where the contributions of the positive correlations between the KLOE08 and KLOE12 uncertainties contribute negatively to the overall uncertainty due to the partial derivative of the $\mu^+\mu^-\gamma$ data in the denominator of the ratio. However, the uncertainties due to a given source are defined in terms of the ratio of $\pi^+\pi^-\gamma$ over ${\mu^+\mu^-\gamma}$, such that the contributions from both data sources are already fully incorporated. Therefore, we do not separately add the uncertainties of these corrections for the $\pi^+\pi^-\gamma$ data to the KLOE12 systematic covariance matrix.

In addition, the KLOE12 systematic uncertainty vector for the non-weighted background subtraction was constructed in~\cite{Babusci:2012rp,KLOE12-KLOEnote} such that it contained the ratio of the contributions from the corrections of the $ee\rightarrow ee \pi\pi$ and $ee\rightarrow ee \mu\mu$ background processes, along with a trackmass ($M_{\rm trk}$) tail correction, summed in quadrature. For this analysis, in order to correctly correlate these independent sources of systematic uncertainty according to equation~\eqref{KLOEerrorprop}, these contributions are separated and correlated individually. This has contributed to the reduction of the KLOE12 error estimate in equation~\eqref{KLOE12}, where previously the correlation of the combined vector resulted in an incorrect overestimate of the systematic uncertainty.

For KLOE08 and KLOE10, the contributions to the systematic uncertainty from the trackmass, tracking efficiency, L3 (software trigger) efficiency, acceptance, luminosity, radiator function, vacuum polarisation correction and final state radiation correction are considered to be fully correlated in the KLOE0810 (KLOE1008) covariance matrix blocks. For the correlation of the systematic uncertainty due to the acceptance, only half of the KLOE10 uncertainty is correlated with the KLOE08 uncertainty in order to ensure that the photon detection acceptance that enters into the KLOE10 uncertainty (that is not present in the KLOE08 analyses) is not correlated and only the correlation of the pion tracks is duly accounted for. Importantly, although the KLOE08 and KLOE10 measurements only overlap for the 50 data points in the energy range 0.35 to 0.85 GeV$^2$, all energy bins in the $60\times75$ KLOE0810 ($75\times60$ KLOE1008) correlation block must be fully correlated. Note that applying 100\% correlation to only the overlapping $50\times50$ region would result in the systematic matrix having negative eigenvalues. 

As with the statistical uncertainties for KLOE0812 (KLOE1208), the systematic uncertainties inherent in the $\pi^+\pi^-\gamma(\gamma)$ data shared between the two analyses are correlated between the KLOE08 and KLOE12 measurements. These include the uncertainties from the L3 efficiency, the background subtraction, the trackmass ($M_{\rm trk}$), the unfolding, the tracking efficiency, the trigger efficiency and the acceptance from the KLOE08 analysis. The determined uncertainties for the L3, $M_{\rm trk}$, tracking, trigger and acceptance corrections for KLOE12 are fully correlated for KLOE0812 such that the anti-correlation that occurs due to the ratio in KLOE12 is propagated accordingly. This is also true for the non-weighted background subtraction contribution, ensuring that only the corrections for the $ee\rightarrow ee \pi\pi$ background from the KLOE08 analysis are correlated with the ratio of the corrections of the $ee\rightarrow ee \pi\pi$ and $ee\rightarrow ee \mu\mu$  background processes as they enter in the KLOE12 analysis. The unfolding uncertainties for the KLOE0812 correlation block are, in part, anti-correlated as they are for KLOE08 and KLOE12 individually. All other systematic uncertainties are 100\% correlated between KLOE08 and KLOE12. 

With the same $\pi^+\pi^-\gamma(\gamma)$ data being shared between the KLOE08 and KLOE12 measurements, the KLOE1012 (KLOE1210) correlation blocks follow a similar structure to the KLOE0810 (KLOE1008) correlation blocks. The caveats to this statement are that there are no correlated uncertainties here due to the luminosity, radiator function or vacuum polarisation correction, as these effects cancel in the ratio of the $\pi^+\pi^-\gamma(\gamma)$ data over the $\mu^+\mu^-\gamma(\gamma)$ data for the KLOE12 measurement (see Section~\ref{Sec:pipixSec}). Therefore, the correlated systematic uncertainties for KLOE1012 are the trackmass, tracking efficiency, L3 efficiency, acceptance and final state radiation correction uncertainties, where it is again necessary to correlate only half of the KLOE10 acceptance uncertainty with KLOE12 in order to ensure that only the effect due to the acceptance of the pion tracks is incorporated.

\section{Combination and results} \label{Sec:KLOEcombination}

\subsection{The combined KLOE $e^+e^-\rightarrow\pi^+\pi^-\gamma(\gamma)$ cross section}

Following the methodology of Section~\ref{Sec:ConstructKLOECovMat} yields full KLOE $\pi^+\pi^-\gamma(\gamma)$ statistical and systematic covariance matrices that describe the correlations that exist between KLOE08, KLOE10 and KLOE12. These data are combined incorporating the energy dependent statistical and systematic uncertainties and corresponding correlations, using an iterative minimisation of the following linear $\chi^2$ function~\cite{Keshavarzi:2018mgv}
\beq \label{IChi^2}
\chi^2 =
\sum^{195}_{i=1}\sum^{195}_{j=1}\big(\sigma^{0}_{\pi\pi(\gamma)}(i)
- \bar{\sigma}^{0}_{\pi\pi(\gamma)}(m)\big) {\bf
  C}^{-1}\big(i^{(m)},j^{(n)}\big)\big(\sigma^{0}_{\pi\pi(\gamma)}(j)
- \bar{\sigma}^{0}_{\pi\pi(\gamma)}(n) \big) \ .
\eeq
Here, $\sigma^{0}_{\pi\pi(\gamma)}(i)$ is the cross section value of the data point $i$
contributing to the combined cross section value $\bar{\sigma}^{0}_{\pi\pi(\gamma)}(m)$ and the combination cross section vector with the elements labelled by $m$ contains 85 data points over the energy range $0.1< s' < 0.95$ GeV$^2$, with the 85 bins corresponding to the 85 distinct energy bins of the three measurements. ${\bf C}^{-1}\big(i^{(m)},j^{(n)}\big)$ is simply the
inverse of the covariance matrix  ${\bf C}\big(i^{(m)},j^{(n)}\big)$, which is defind as
the sum of the statistical covariance matrix ${\rm C}^{\rm
  stat}\big(i^{(m)},j^{(n)}\big)$ and the systematic covariance matrix
${\rm C}^{\rm sys}\big(i^{(m)},j^{(n)}\big)$. At each iterative stage of the
minimisation, it is defined as 
\beq \label{ICk}
{\bf C}\big(i^{(m)},j^{(n)}\big) = {\text
  C}^{\text{stat}}\big(i^{(m)},j^{(n)}\big) + \frac{{\text
    C}^{\text{sys}}\big(i^{(m)},j^{,n)}\big)}{\sigma^{0}_{\pi\pi(\gamma)}(i)\sigma^{0}_{\pi\pi(\gamma)}(j)}
\bar{\sigma}^{0}_{\pi\pi(\gamma)}(m)\bar{\sigma}^{0}_{\pi\pi(\gamma)}(n) \,, 
\eeq
where the quantities $\bar{\sigma}^{0}_{\pi\pi(\gamma)}(m)$ and
$\bar{\sigma}^{0}_{\pi\pi(\gamma)}(n)$ are the resulting combined cross section values from the previous iteration. This method has been adapted from~\cite{Ball:2009qv} (see also~\cite{Benayoun:2015gxa}), has been advocated to be free of systematic bias and exhibits a swift convergence, after only a few iterations. 
\begin{table}[!h]
\small
\hspace{0.8cm}
  \begin{tabular}{|c|c|c|c|c|c|}
  \hline 
  \multicolumn{6}{|c|}{KLOE combination} \\
  \hline
 $s'({\rm GeV^2})$   & $\sigma^{0}_{\pi\pi(\gamma)}({\rm nb})$ & $ \abs{F(\pi)}^2$ & $s'({\rm GeV^2})$   & $\sigma^{0}_{\pi\pi(\gamma)}({\rm nb})$ & $ \abs{F(\pi)}^2$  \\ 
  \hline
0.105	&	47.27	$\pm$	8.41	&	      1.74	$\pm$	      0.31	&	0.535	&	1154.56	$\pm$	6.81	&	     35.96	$\pm$	      0.21	\\
0.115	&	70.65	$\pm$	10.44	&	      2.04	$\pm$	      0.30	&	0.545	&	1207.69	$\pm$	6.83	&	     38.20	$\pm$	      0.22	\\
0.125	&	80.13	$\pm$	10.97	&	      2.00	$\pm$	      0.27	&	0.555	&	1243.32	$\pm$	10.13	&	     39.94	$\pm$	      0.33	\\
0.135	&	80.42	$\pm$	11.27	&	      1.82	$\pm$	      0.26	&	0.565	&	1285.35	$\pm$	7.14	&	     41.92	$\pm$	      0.23	\\
0.145	&	87.58	$\pm$	11.70	&	      1.86	$\pm$	      0.25	&	0.575	&	1277.36	$\pm$	7.32	&	     42.29	$\pm$	      0.24	\\
0.155	&	102.88	$\pm$	12.35	&	      2.10	$\pm$	      0.25	&	0.585	&	1279.89	$\pm$	7.31	&	     42.98	$\pm$	      0.25	\\
0.165	&	115.16	$\pm$	13.85	&	      2.29	$\pm$	      0.28	&	0.595	&	1274.03	$\pm$	10.32	&	     43.27	$\pm$	      0.35	\\
0.175	&	122.58	$\pm$	13.42	&	      2.40	$\pm$	      0.26	&	0.605	&	1228.97	$\pm$	12.29	&	     42.18	$\pm$	      0.42	\\
0.185	&	126.19	$\pm$	12.61	&	      2.45	$\pm$	      0.24	&	0.615	&	950.47	$\pm$	20.95	&	     34.85	$\pm$	      0.77	\\
0.195	&	146.34	$\pm$	14.10	&	      2.84	$\pm$	      0.27	&	0.625	&	803.87	$\pm$	4.65	&	     29.94	$\pm$	      0.17	\\
0.205	&	144.18	$\pm$	13.35	&	      2.80	$\pm$	      0.26	&	0.635	&	781.82	$\pm$	4.39	&	     29.24	$\pm$	      0.16	\\
0.215	&	147.47	$\pm$	12.68	&	      2.88	$\pm$	      0.25	&	0.645	&	731.86	$\pm$	5.74	&	     27.61	$\pm$	      0.22	\\
0.225	&	154.64	$\pm$	11.98	&	      3.04	$\pm$	      0.24	&	0.655	&	679.26	$\pm$	3.93	&	     25.90	$\pm$	      0.15	\\
0.235	&	170.47	$\pm$	12.40	&	      3.39	$\pm$	      0.25	&	0.665	&	620.73	$\pm$	3.46	&	     23.93	$\pm$	      0.13	\\
0.245	&	168.96	$\pm$	11.53	&	      3.40	$\pm$	      0.23	&	0.675	&	569.26	$\pm$	4.63	&	     22.20	$\pm$	      0.18	\\
0.255	&	176.55	$\pm$	10.84	&	      3.60	$\pm$	      0.22	&	0.685	&	518.39	$\pm$	5.62	&	     20.45	$\pm$	      0.22	\\
0.265	&	202.38	$\pm$	11.63	&	      4.18	$\pm$	      0.24	&	0.695	&	471.79	$\pm$	2.69	&	     18.82	$\pm$	      0.11	\\
0.275	&	203.28	$\pm$	10.70	&	      4.26	$\pm$	      0.22	&	0.705	&	431.19	$\pm$	2.44	&	     17.39	$\pm$	      0.10	\\
0.285	&	215.28	$\pm$	10.60	&	      4.58	$\pm$	      0.23	&	0.715	&	386.51	$\pm$	3.21	&	     15.76	$\pm$	      0.13	\\
0.295	&	225.63	$\pm$	10.46	&	      4.87	$\pm$	      0.23	&	0.725	&	356.81	$\pm$	2.03	&	     14.70	$\pm$	      0.08	\\
0.305	&	236.90	$\pm$	10.49	&	      5.19	$\pm$	      0.23	&	0.735	&	327.36	$\pm$	1.91	&	     13.63	$\pm$	      0.08	\\
0.315	&	244.65	$\pm$	10.11	&	      5.45	$\pm$	      0.23	&	0.745	&	299.08	$\pm$	1.96	&	     12.59	$\pm$	      0.08	\\
0.325	&	248.45	$\pm$	9.83	&	      5.62	$\pm$	      0.22	&	0.755	&	273.28	$\pm$	1.80	&	     11.62	$\pm$	      0.08	\\
0.335	&	255.64	$\pm$	9.62	&	      5.88	$\pm$	      0.22	&	0.765	&	249.34	$\pm$	1.45	&	     10.71	$\pm$	      0.06	\\
0.345	&	280.05	$\pm$	9.46	&	      6.54	$\pm$	      0.22	&	0.775	&	228.91	$\pm$	1.94	&	      9.93	$\pm$	      0.08	\\
0.355	&	305.24	$\pm$	4.55	&	      7.24	$\pm$	      0.11	&	0.785	&	211.31	$\pm$	1.27	&	      9.26	$\pm$	      0.06	\\
0.365	&	330.21	$\pm$	7.67	&	      7.96	$\pm$	      0.18	&	0.795	&	196.17	$\pm$	1.36	&	      8.68	$\pm$	      0.06	\\
0.375	&	349.58	$\pm$	4.60	&	      8.56	$\pm$	      0.11	&	0.805	&	183.29	$\pm$	1.08	&	      8.19	$\pm$	      0.05	\\
0.385	&	376.70	$\pm$	4.63	&	      9.37	$\pm$	      0.12	&	0.815	&	170.45	$\pm$	1.00	&	      7.69	$\pm$	      0.05	\\
0.395	&	400.82	$\pm$	4.57	&	     10.12	$\pm$	      0.12	&	0.825	&	157.72	$\pm$	1.09	&	      7.19	$\pm$	      0.05	\\
0.405	&	433.99	$\pm$	6.28	&	     11.13	$\pm$	      0.16	&	0.835	&	146.52	$\pm$	0.95	&	      6.74	$\pm$	      0.04	\\
0.415	&	465.70	$\pm$	4.79	&	     12.13	$\pm$	      0.12	&	0.845	&	136.86	$\pm$	0.79	&	      6.36	$\pm$	      0.04	\\
0.425	&	506.53	$\pm$	4.87	&	     13.39	$\pm$	      0.13	&	0.855	&	126.97	$\pm$	0.78	&	      5.95	$\pm$	      0.04	\\
0.435	&	544.42	$\pm$	4.84	&	     14.61	$\pm$	      0.13	&	0.865	&	119.05	$\pm$	0.89	&	      5.63	$\pm$	      0.04	\\
0.445	&	585.65	$\pm$	5.04	&	     15.95	$\pm$	      0.14	&	0.875	&	111.33	$\pm$	0.83	&	      5.31	$\pm$	      0.04	\\
0.455	&	640.09	$\pm$	7.95	&	     17.69	$\pm$	      0.22	&	0.885	&	104.92	$\pm$	1.81	&	      5.05	$\pm$	      0.09	\\
0.465	&	691.86	$\pm$	7.66	&	     19.41	$\pm$	      0.21	&	0.895	&	98.60	$\pm$	0.59	&	      4.79	$\pm$	      0.03	\\
0.475	&	740.82	$\pm$	8.20	&	     21.09	$\pm$	      0.23	&	0.905	&	93.05	$\pm$	0.56	&	      4.56	$\pm$	      0.03	\\
0.485	&	822.23	$\pm$	5.82	&	     23.75	$\pm$	      0.17	&	0.915	&	87.66	$\pm$	0.74	&	      4.33	$\pm$	      0.04	\\
0.495	&	895.61	$\pm$	17.85	&	     26.26	$\pm$	      0.52	&	0.925	&	82.76	$\pm$	0.49	&	      4.13	$\pm$	      0.02	\\
0.505	&	953.15	$\pm$	13.08	&	     28.36	$\pm$	      0.39	&	0.935	&	78.84	$\pm$	0.65	&	      3.96	$\pm$	      0.03	\\
0.515	&	1032.72	$\pm$	6.28	&	     31.20	$\pm$	      0.19	&	0.945	&	74.74	$\pm$	0.64	&	      3.79	$\pm$	      0.03	\\
0.525	&	1078.01	$\pm$	8.23	&	     33.06	$\pm$	      0.25	&	-	&			-	&			-	\\
\hline
  \end{tabular}
    \caption{\small The combined KLOE measurement of the $\pi^+\pi^-\gamma(\gamma)$ bare cross section and pion form factor in $0.01$ GeV$^2$ intervals from $0.10< s' < 0.95$ GeV$^2$. Here, $s'$ denotes the bin centre. For both $\sigma^{0}_{\pi\pi(\gamma)}$ and $\abs{F(\pi)}^2$, the error shown is the total (statistical and systematic) uncertainty. The errors have been inflated according to the local $\chi^2_{\rm min}/{\rm d.o.f.}$ in each energy bin, where inflation is only applied if $\chi^2_{\rm min}/{\rm d.o.f.}>1$.}\label{Tab:KLOEcombination}
    \end{table} 
We also obtain an output covariance matrix for the combined statistical and systematic uncertainties that describes the correlations between the data points of the resulting cross section vector. 

The KLOE combination cross section and pion form factor data are listed in Table~\ref{Tab:KLOEcombination}. The input cross section vectors and combination covariance matrices, along with the combined output cross section vector and total covariance matrix are available from~\cite{KLOE17ppg}.\footnote{The total output matrix given contains the contributions from both the statistical and systematic uncertainties, where the choice to use both as input into the data combination results in a solution that entangles the statistical and systematic sources of uncertainty.}
 \begin{figure}[htbp]
\centering
\vspace{-1cm}
\hspace{-0.5cm}
 {\subfloat[Cross section in the full data range]{%
    \includegraphics[width= 0.48\textwidth]{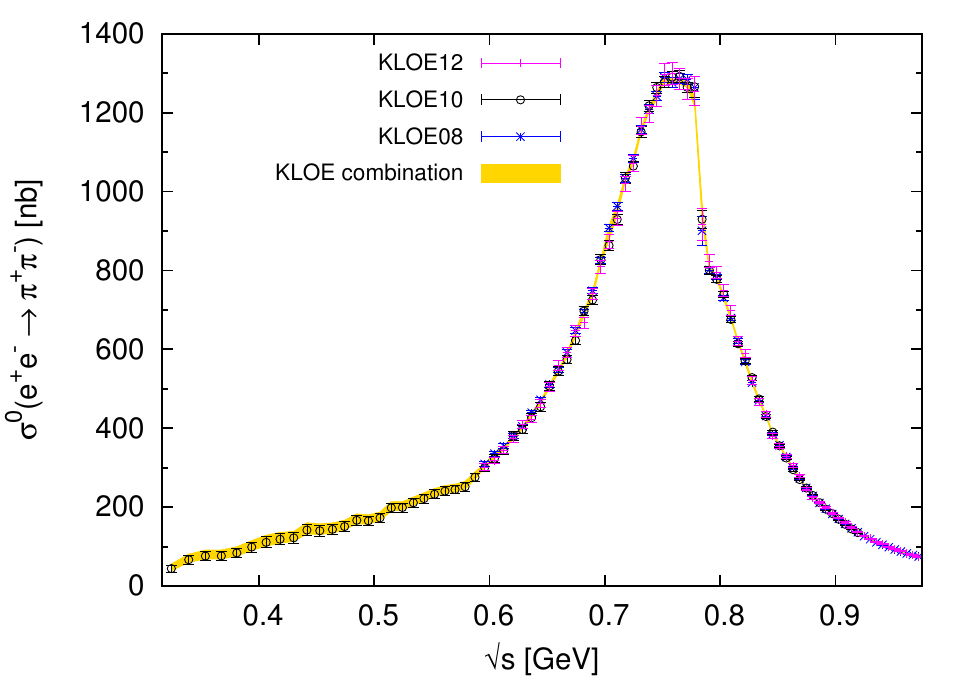}}    \hspace{0.4cm} 
  \subfloat[Cross section in the overlapping data range]{%
    \includegraphics[width= 0.48\textwidth]{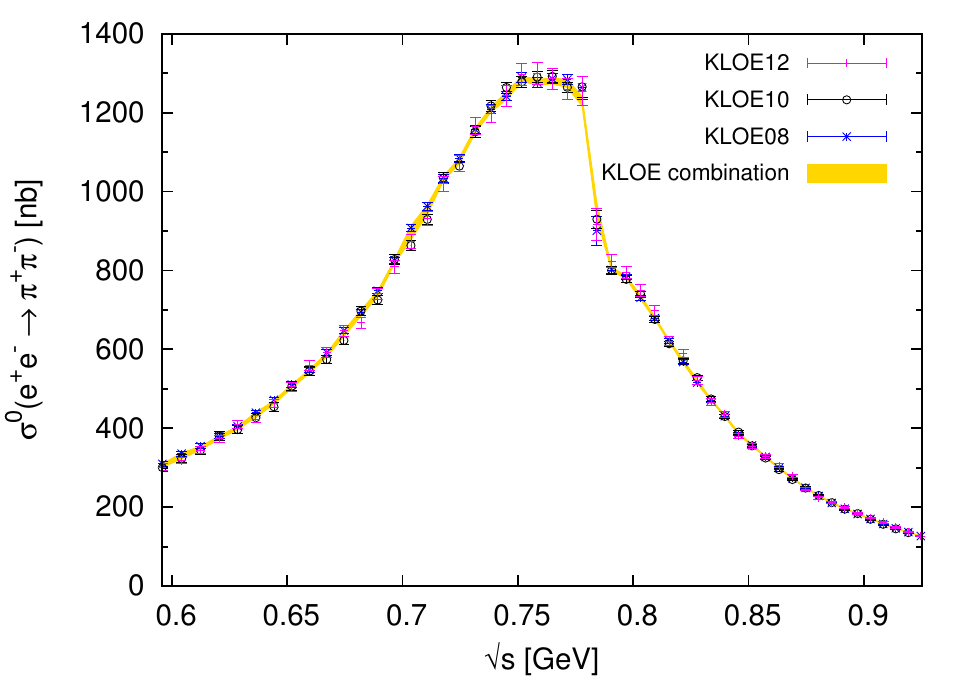}}\hfill 
  \caption{\small The KLOE combination plotted with the individual cross section measurements, where the KLOE combination is represented by the yellow band and the KLOE08, KLOE10 and KLOE12 cross section measurements are given by the blue, black and pink markers, respectively (colour online). In all cases, the error bars shown are the statistical and systematic uncertainties summed in quadrature.}\label{KLOEcombination} }
    {\vspace{-0.25cm}\subfloat[Normalised difference in the full data range]{%
    \includegraphics[width= 0.5\textwidth]{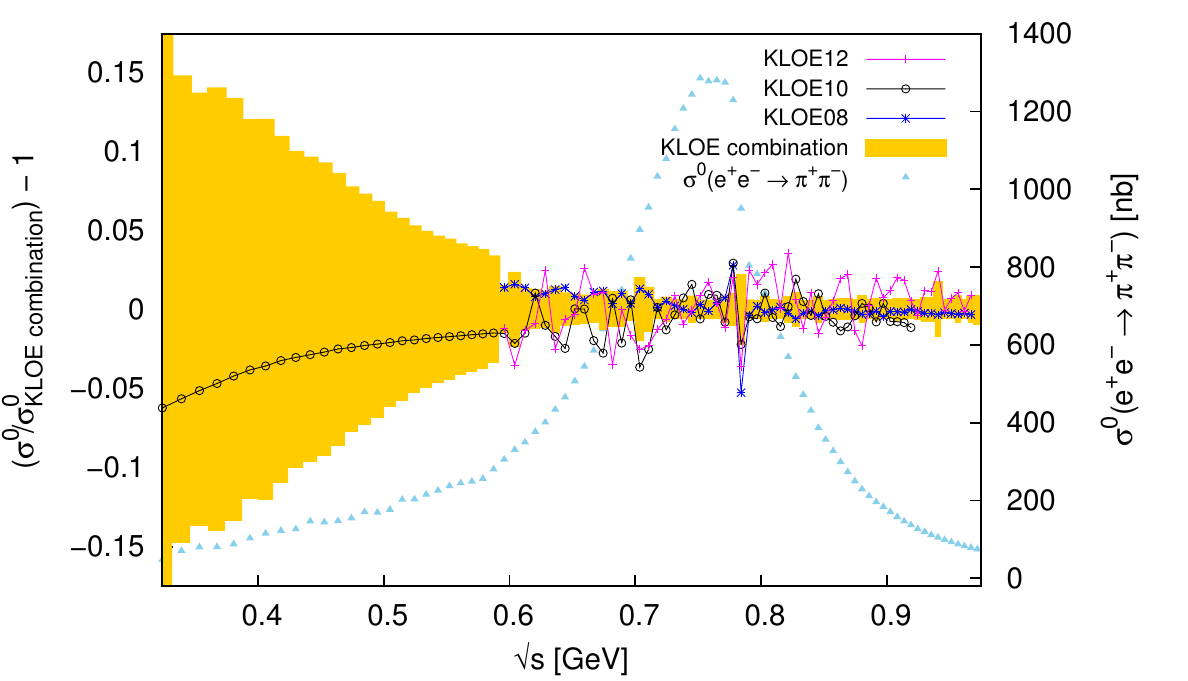}}\hfill
  \subfloat[Normalised difference in the overlapping data range]{%
    \includegraphics[width= 0.5\textwidth]{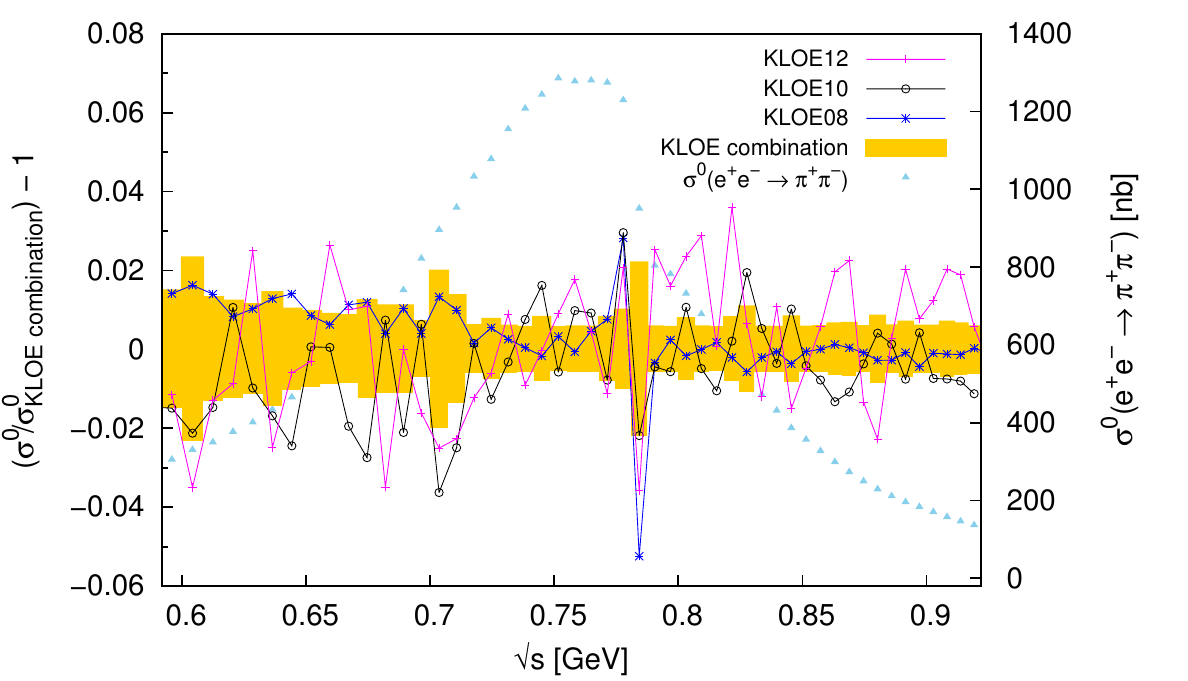}}\hfill 
  \caption{\small The normalised difference of the KLOE combination and the individual KLOE measurements, where the yellow band represents the statistical and systematic uncertainties of the KLOE combination summed in quadrature and the KLOE08, KLOE10 and KLOE12 cross section measurements are given by the blue, black and pink markers, respectively (colour online). Here, the errors bars of the individual measurements are not shown in order to be able to distinguish the data points, but are in good agreement with the KLOE combination.}\label{KLOEcombinationDiff} } 
{\vspace{0.cm}{%
    \includegraphics[width= 0.55\textwidth]{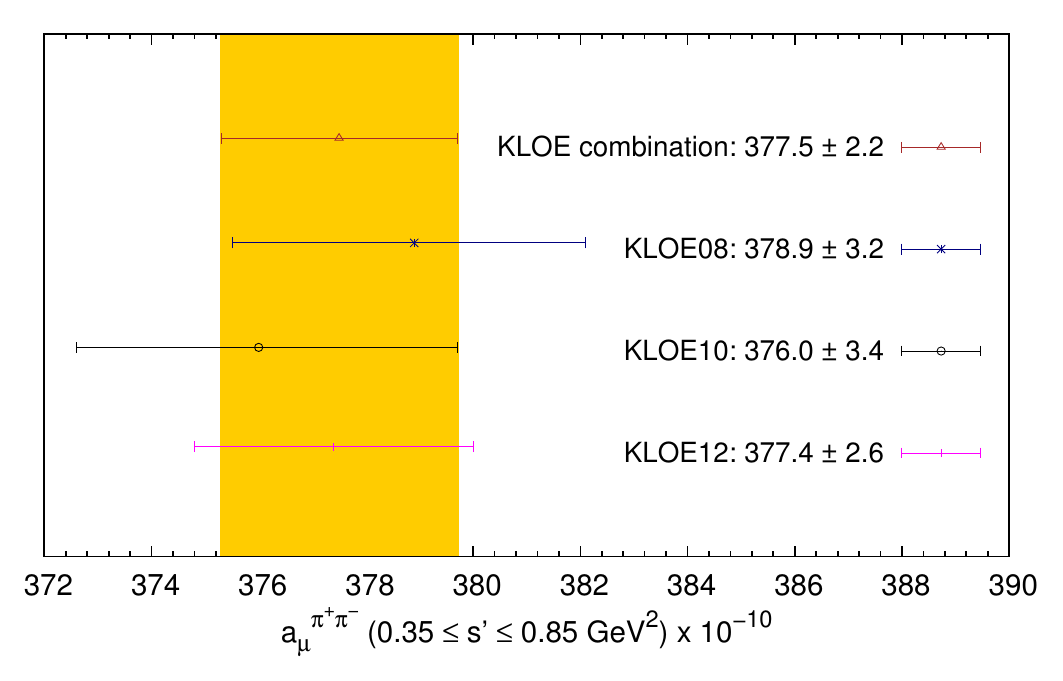}} \hfill
  \caption{\small Comparison of estimates of $a_{\mu}^{\pi^+\pi^-}$ from the KLOE combination and the individual KLOE measurements in the range $0.35 < s' < 0.85$ GeV$^2$. The KLOE combination is represented by the yellow band (colour online). In all cases, the uncertainties shown are the statistical and systematic uncertainties summed in quadrature.}  \label{KLOEamuCompare}}
\end{figure} 
 \begin{figure}[t]
\centering
  {%
  \includegraphics[width= 0.7\textwidth]{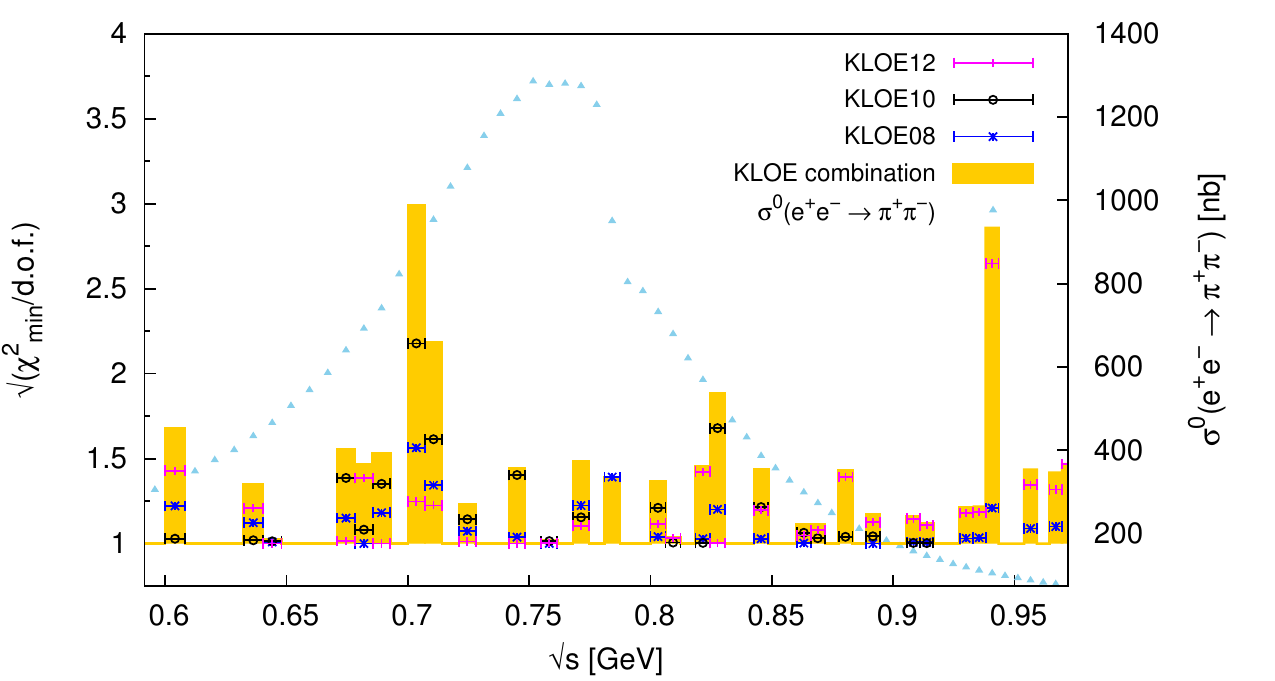} \hfill
  \caption{\small The effect of the local $\chi^2_{\rm min}/{\rm d.o.f.}$ error inflation on the KLOE combination, which is applied in each energy bin if the local $\chi^2_{\rm min}/{\rm d.o.f.}>1$. The total effect on the KLOE combination is represented by the yellow blocks (colour online). The relative contributions to each local  $\chi^2_{\rm min}/{\rm d.o.f.}$ from the KLOE08, KLOE10 and KLOE12 measurements individually are given by the blue, black and pink markers, respectively.}  \label{KLOEchi2}}
\end{figure} 
\begin{table}[t]
\centering
  \begin{tabular}{|l|c|}
  \hline 
 KLOE $\pi^+\pi^-\gamma(\gamma)$ data set  & $a_{\mu}^{\pi^+\pi^-} (0.35< s' < 0.85 \ {\rm GeV}^2)$  \\ 
  \hline
  KLOE08 & $378.9 \pm 0.4_{\rm stat} \pm 3.2_{\rm sys}$  \\
  KLOE10 & $376.0 \pm 0.9_{\rm stat} \pm 3.3_{\rm sys}$   \\
  KLOE12 & $377.4 \pm 1.2_{\rm stat} \pm 2.3_{\rm sys}$   \\
\hline
  KLOE combination & $377.5 \pm 0.5_{\rm stat} \pm 2.1_{\rm sys}$ \\
  \hline
  \end{tabular}
    \caption{\small Comparative results of the values obtained for $a_{\mu}^{\pi^+\pi^-}(0.35< s' < 0.85 \ {\rm GeV}^2)$ from the individual KLOE measurements and the full combination. Results for $a_{\mu}^{\pi^+\pi^-}$ are given in units of $10^{-10}$.}\label{Tab:amu}
    \end{table} 
For the contribution to the anomalous magnetic moment of the muon in the full energy range, the KLOE combination results in
\beq \label{KLOEcombination}
a_{\mu}^{\pi^+\pi^-}(0.10< s' <0.95 \text{ GeV}^2) = (489.8 \pm 1.7_{\rm stat} \pm 4.8_{\rm sys} ) \times 10^{-10} .
\eeq
The resulting cross section and the individual measurements are shown in Figure~\ref{KLOEcombination}. In addition, the normalised differences of the individual KLOE measurements and the combination are shown in Figure~\ref{KLOEcombinationDiff}. We observe good agreement between the data and the combination, especially with KLOE08 which dominates the fit due to its smaller statistical uncertainty when comparing to KLOE10 and KLOE12. KLOE12 exhibits the largest fluctuations when comparing to the fitted combination, but is well within the errors of the data. In plot (a) of Figure~\ref{KLOEcombinationDiff}, we note how the KLOE0810 and KLOE1012 systematic uncertainties have a non-trivial effect in the lower energy region where only the KLOE10 data exist, with the correlations providing an expected upward pull (which is well within the errors of the combination) to the KLOE combination cross section away from the KLOE10 data points. 

For the overlapping energy region of all three measurements, the estimates for $a_{\mu}^{\pi^+\pi^-}$ from the KLOE combination and the individual measurements are given in Table~\ref{Tab:amu} and Figure~\ref{KLOEamuCompare}. In all cases, the errors include all correlation contributions. For the combination, they have been inflated according to a local $\chi^2_{\rm min}/{\rm d.o.f.}$ in each energy bin if the $\chi^2_{\rm min}/{\rm d.o.f.}>1$~\cite{PDG2016,Eidelman:1995ny,Davier:2009zi}, as shown in Figure~\ref{KLOEchi2}. This has resulted in an increase to the overall uncertainty of the estimate of $a_{\mu}^{\pi^+\pi^-}$ of $\sim13\%$. The combination agrees well with the estimates from the individual measurements, with a marked improvement in the overall uncertainty. While the statistical uncertainty of $a_\mu^{\pi^+\pi^-}$ from the combination is dominated by KLOE08 (which has the smallest statistical uncertainty of the three individual measurements), the combination mean value of $a_{\mu}^{\pi^+\pi^-}$ is closest to that obtained with the KLOE12 data alone, which has the smallest systematic and, therefore, the smallest total error of the three. This in turn leads to the improved systematic error of the combined result and its markedly improved total error.

\subsection{Comparison with results from the CMD-2, SND, BaBar and BESIII experiments}

 \begin{figure}[t]
\centering
  {\subfloat[Cross section in the range $0.6 < \sqrt{s'} < 0.9$ GeV]{%
    \includegraphics[width= 0.5\textwidth]{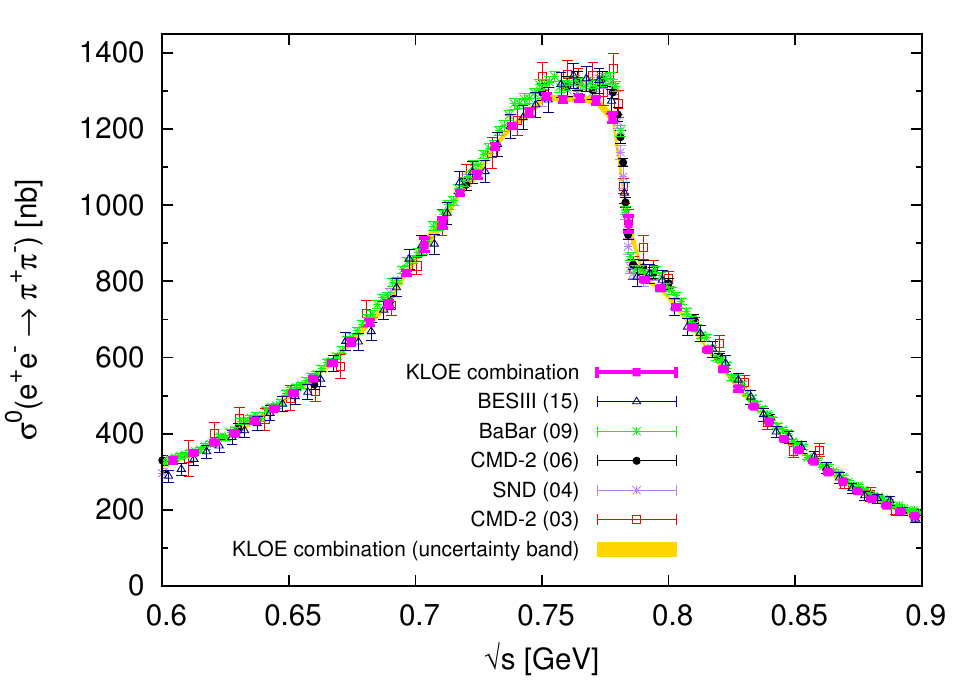}} \hfill
        \subfloat[Cross section in the $\rho-\omega$ interference region ]{%
    \includegraphics[width= 0.5\textwidth]{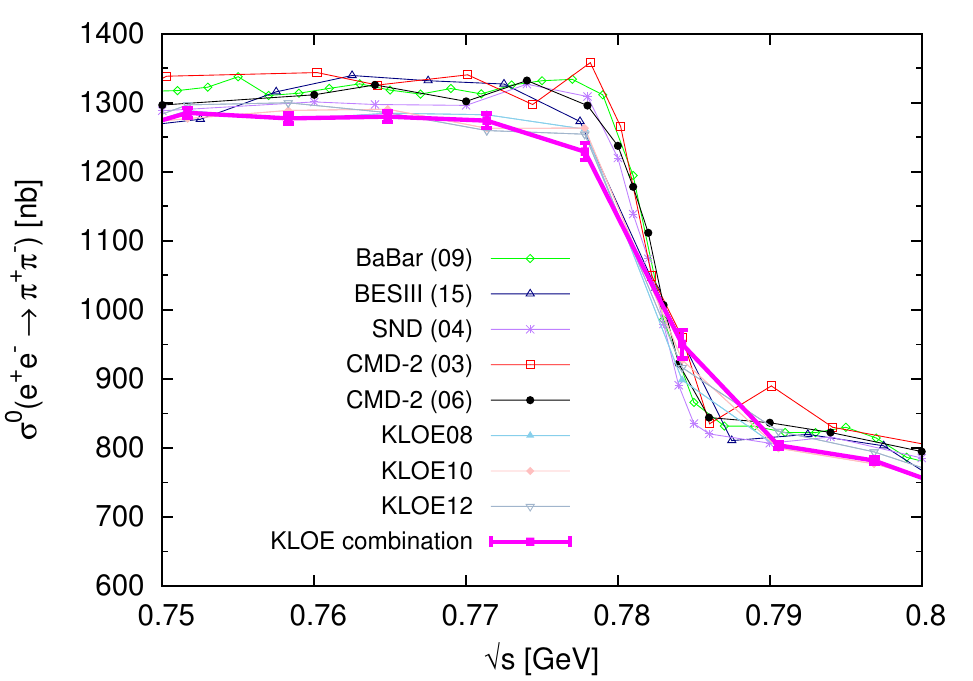}} \hfill
\caption{\small The $\pi^+\pi^-$ cross section from the KLOE combination, CMD-2~\cite{Akhmetshin:2006wh,Akhmetshin:2006bx,Akhmetshin:2003zn}, SND~\cite{Achasov:2006vp}, BaBar~\cite{Aubert:2009ad} and BESIII~\cite{Ablikim:2015orh} data points. The KLOE combination is represented by the yellow band (colour online). Where uncertainties are displayed, they represent the statistical and systematic uncertainties summed in quadrature. The uncertainties of the separate experimental measurements in Figure (b) have been suppressed in order to improve readability.}  \label{ExpxSecCompare}}
\end{figure} 

 \begin{figure}[htbp]
\centering
  \subfloat[KLOE combination vs. other experiments]{%
    \includegraphics[width= 0.6\textwidth]{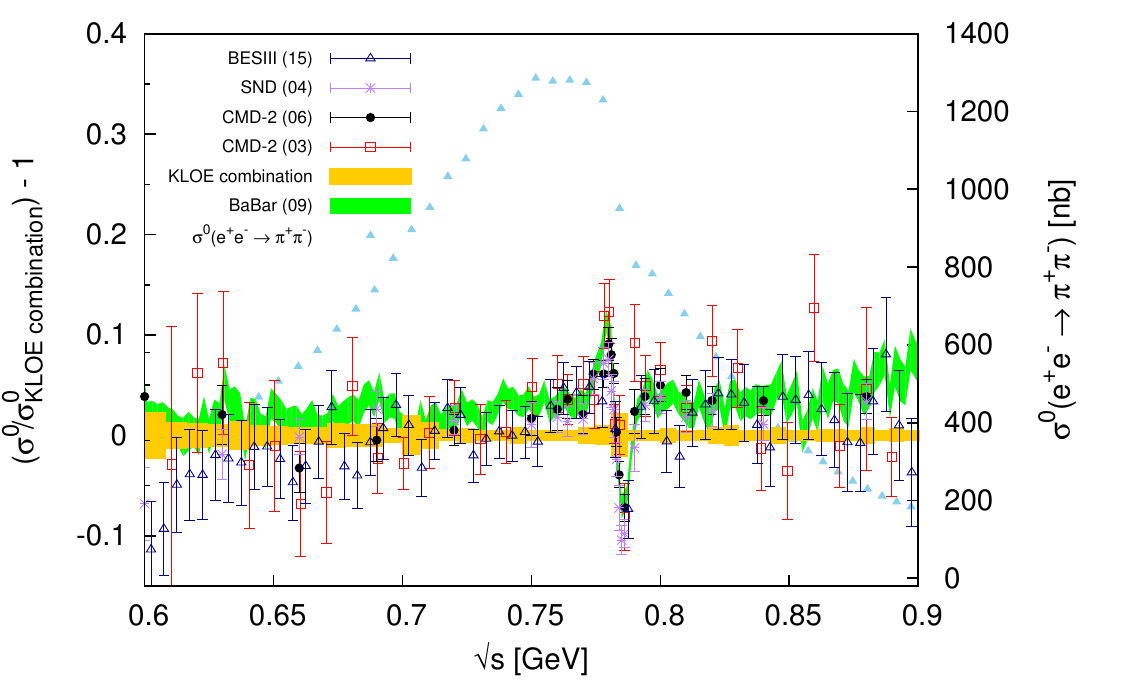}} \hfill
 \subfloat[KLOE combination vs. BaBar]{%
    \includegraphics[width= 0.5\textwidth]{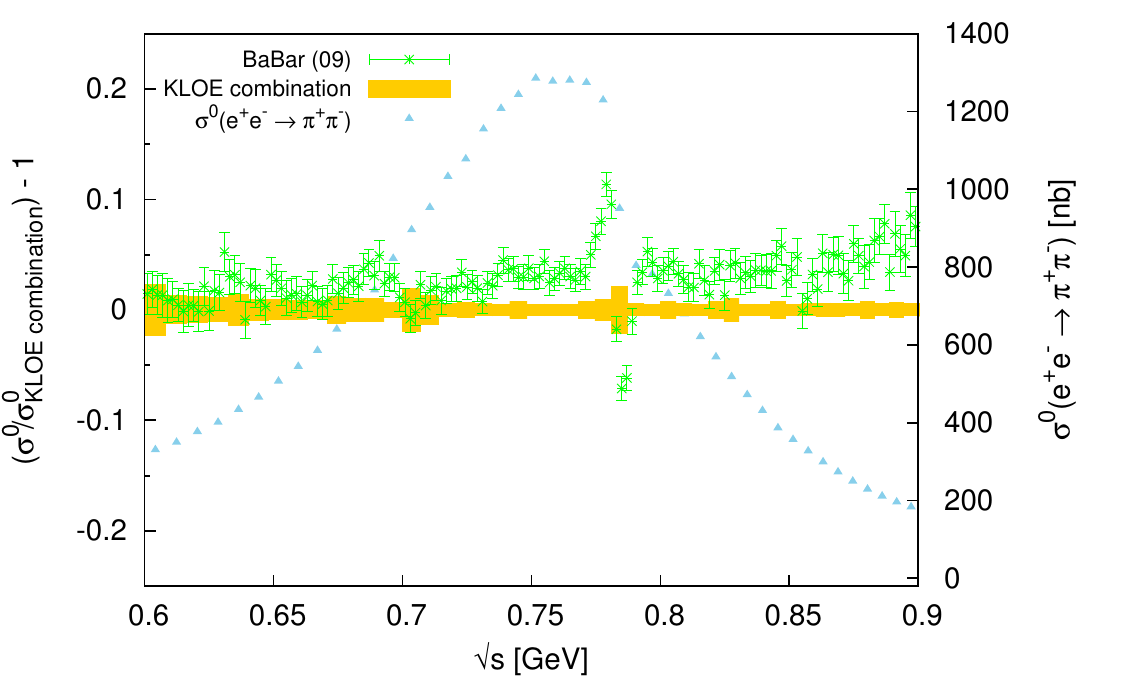}}\hfill
        \subfloat[KLOE combination vs. BESIII]{%
    \includegraphics[width= 0.5\textwidth]{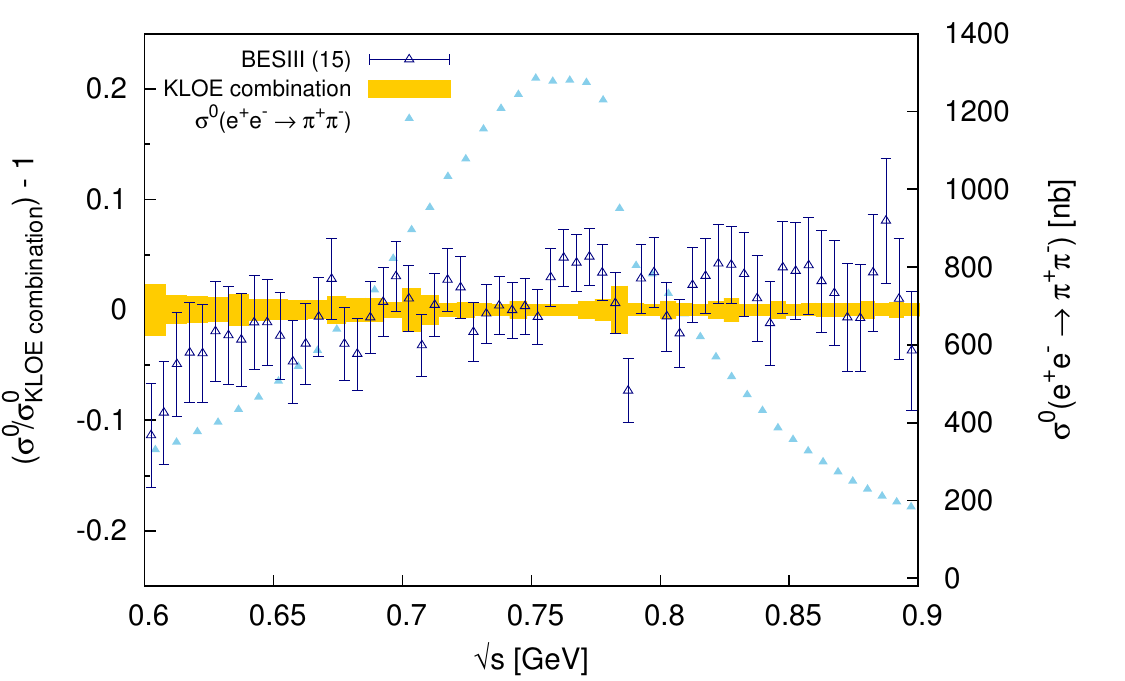}}\hfill
     \subfloat[KLOE combination vs. CMD-2]{%
    \includegraphics[width= 0.5\textwidth]{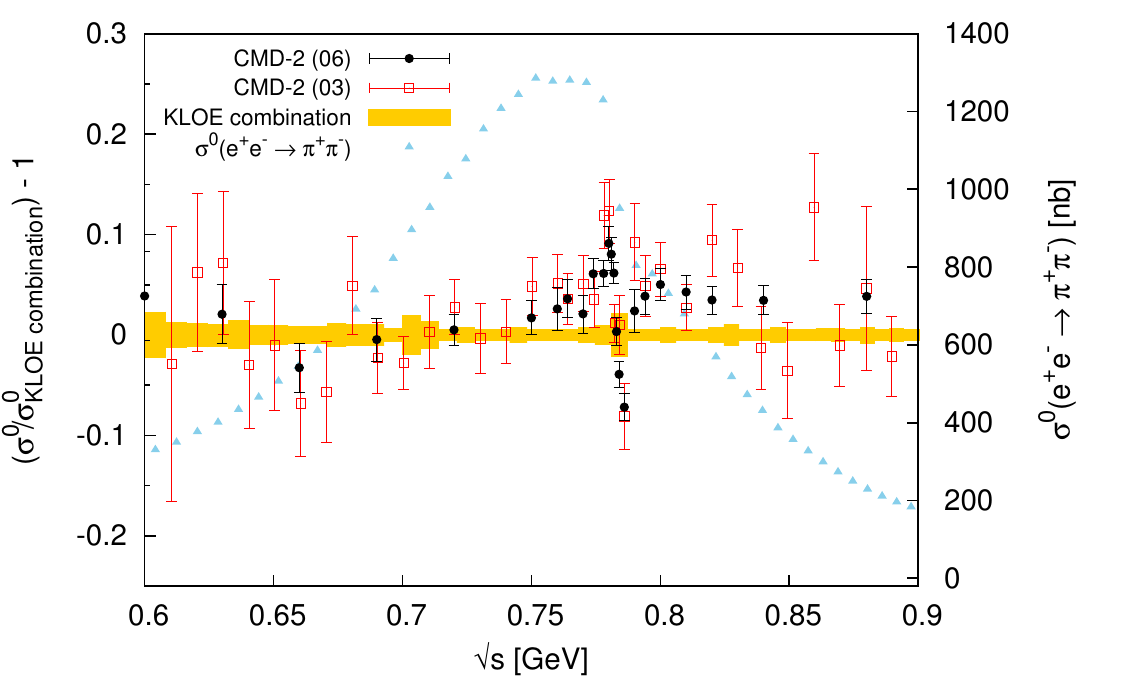}}\hfill
 \subfloat[KLOE combination vs. SND]{%
    \includegraphics[width= 0.5\textwidth]{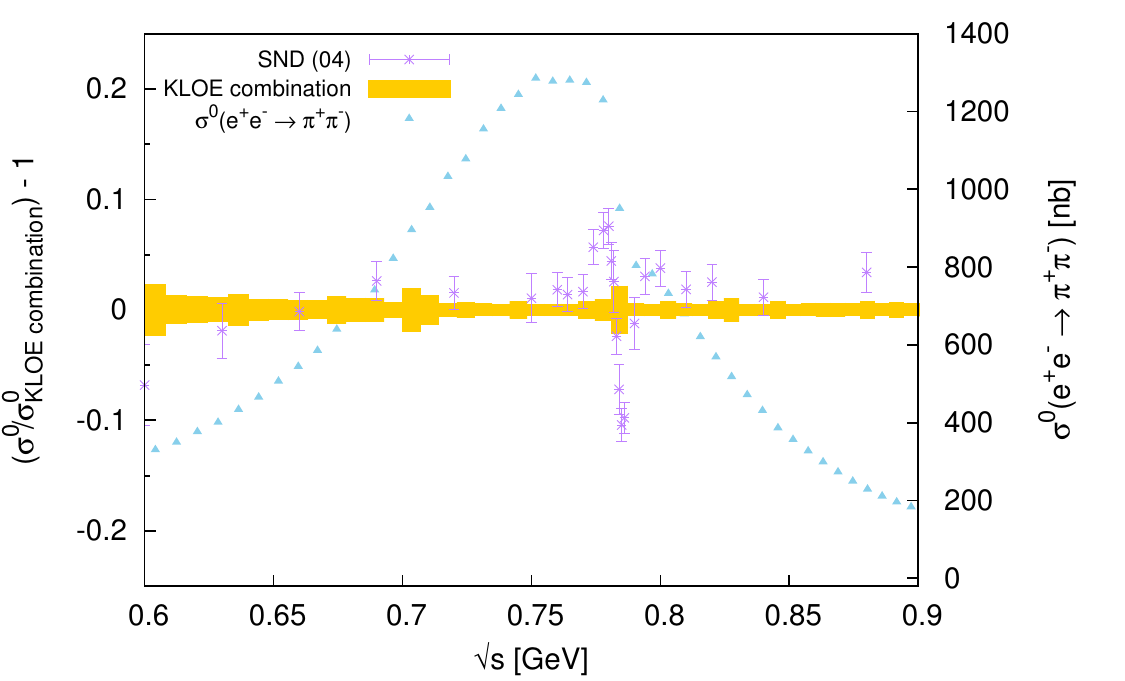}}\hfill
  \caption{\small The $\pi^+\pi^-$ cross section from the KLOE combination compared to the CMD-2, SND, BaBar and BESIII data points in the range $0.6 < \sqrt{s'} < 0.9$ GeV. The KLOE combination is represented by the yellow band  (colour online). In all cases, the uncertainties shown are the statistical and systematic uncertainties summed in quadrature.}  \label{ExpCompareDiff}
\end{figure} 

 \begin{figure}[htbp]
\centering
  {%
    \includegraphics[width= 0.6\textwidth]{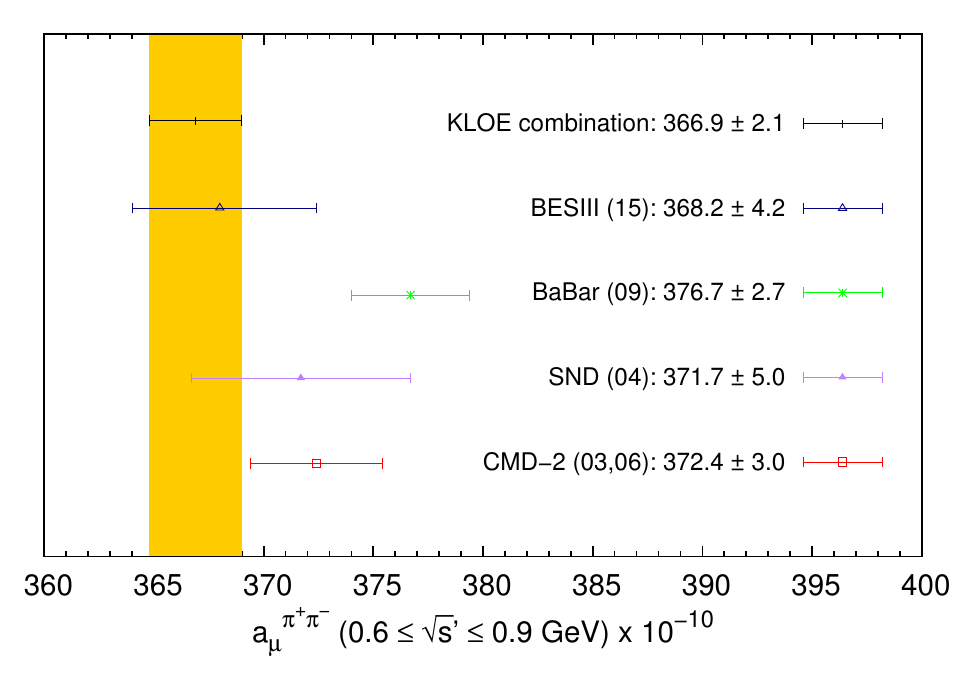}    \hspace{-0.5cm} \hfill
  \caption{\small Estimates of $a_{\mu}^{\pi^+\pi^-}$ from the KLOE combination, CMD-2, SND, BaBar and BESIII in the range $0.6 < \sqrt{s'} < 0.9$ GeV. The available CMD-2 data have been combined following the prescription of~\cite{Keshavarzi:2018mgv}. The KLOE combination is represented by the yellow band (colour online). In all cases, the uncertainties shown are the statistical and systematic uncertainties summed in quadrature.}  \label{ExpamuCompare}}
\end{figure} 

The $\sigma(e^+e^-\rightarrow\pi^+\pi^-)$ cross section has been measured below $1$ GeV by the CMD-2~\cite{Akhmetshin:2006wh,Akhmetshin:2006bx,Akhmetshin:2003zn}, SND~\cite{Achasov:2006vp}, BaBar~\cite{Aubert:2009ad} and BESIII~\cite{Ablikim:2015orh} collaborations. The BaBar and BESIII measurements, like the KLOE measurements, are obtained through radiative return. The CMD-2 and SND measurements are taken by energy scan, allowing us to compare the two methods. All the experimental measurements are undressed of VP effects and include FSR, such that there is a consistent comparison of $\sigma^0_{\pi\pi(\gamma)}$. The cross section measurements from each experiment and the KLOE combination are shown in Figure~\ref{ExpxSecCompare}.

The normalised difference of the data from these experiments with respect to the KLOE combination are shown in Figure~\ref{ExpCompareDiff}. In particular, we note that the KLOE combination is lower than all other data at the $\rho$ peak where the cross section is largest, but higher than the other experimental data where the cross section drops off in the $\rho-\omega$ interference region. This effect is evident in Figure~\ref{ExpCompareDiff}, where we note that for all cases (except for (c) BESIII, where the effect is less prominent), there is a sharp rise and fall of the difference in the experimental cross section at the $\rho-\omega$ interference region due to KLOE having fewer bins in this region compared to the other experiments (see plot (b) of Figure~\ref{ExpxSecCompare}). 

The BaBar data are, in majority, higher than the KLOE combination, whereas we observe that the other data sit mainly lower than KLOE below the $\rho$ peak and higher above it. We also note that our comparison of the KLOE combination with the BESIII data looks markedly different from that presented in~\cite{Ablikim:2015orh}, especially at higher energies. However, in~\cite{Ablikim:2015orh}, the comparison has been made using a fit of the data to the Gounaris-Sakurai parametrisation~\cite{Gounaris:1968mw}, which does not provide an adequate description of the BESIII measurements of the $\pi^+\pi^-$ cross section in the tail of the resonance. We therefore opt to compare, in plot (c) of Figure~\ref{ExpCompareDiff}, the published BESIII data points directly with our combination of the KLOE data.

\begin{table}[!t]
\centering
  \begin{tabular}{|l|c|}
  \hline 
 $\pi^+\pi^-\gamma(\gamma)$ data set  & $a_{\mu}^{\pi^+\pi^-} (0.6< \sqrt{s'} < 0.9 \ {\rm GeV})$  \\ 
  \hline
  CMD-2 fit (03,06) & $372.4 \pm 3.0$   \\
  SND (04) & $371.7 \pm 5.0$   \\
  BaBar (09) & $376.7 \pm 2.7$   \\
  BESIII (15) & $368.2 \pm 4.2$  \\
\hline
  KLOE combination & $366.9 \pm 2.1$ \\
  \hline
  \end{tabular}
    \caption{\small Comparative results of the values obtained for $a_{\mu}^{\pi^+\pi^-}(0.6< \sqrt{s'} < 0.9 \ {\rm GeV})$ from the KLOE combination and the CMD-2, SND, BaBar and BESIII data. The available CMD-2 data have been combined following the prescription of~\cite{Keshavarzi:2018mgv}. Results for $a_{\mu}^{\pi^+\pi^-}$ are given in units of $10^{-10}$. In all cases, the uncertainties shown are the statistical and systematic uncertainties summed in quadrature.}\label{Tab:ExpCompare}
\end{table} 

Estimates of the contribution to the anomalous magnetic moment of the muon from these experiments in the range $0.6 < \sqrt{s'} < 0.9$ GeV are shown in Figure~\ref{ExpamuCompare} and Table~\ref{Tab:ExpCompare}, where we have combined the available CMD-2 data into a single estimate by applying the same method used to fit the KLOE combination. We observe good agreement (within $1.5\sigma$) between the KLOE combination and the measurements by CMD-2, SND and BESIII. The measurement by BaBar, as evident from plot (b) of Figure~\ref{ExpCompareDiff}, results in a higher estimate of $a_\mu^{\pi^+\pi^-}$.

\section{Conclusions}

The KLOE collaboration have performed three measurements of the $\sigma^0\big(e^+e^-\allowbreak\rightarrow\pi^+\pi^-\gamma(\gamma)\big)$ cross section below $1$ GeV$^2$ using the method of radiative return. These measurements are, in part, highly correlated. This is especially true for KLOE08 and KLOE12 where, for the KLOE12 measurement, the KLOE08 $\pi^+\pi^-\gamma(\gamma)$ data is normalised by the measured $\mu^+\mu^-\gamma(\gamma)$ cross section. This has necessitated the construction of statistical and systematic combination covariance matrices, which have been carefully built to satisfy the required properties of a covariance matrix. 

Using these covariance matrices, the three measurements have been combined to produce single vectors for both the two-pion cross section $\sigma_{\pi\pi(\gamma)}$ and the pion form factor $\left|F_{\pi}\right|^2$, along with a corresponding covariance matrix for each. This combination of the KLOE cross section data results in an estimate of the two-pion contribution to the anomalous magnetic moment of the muon of
\beq
a_{\mu}^{\pi^+\pi^-}({\rm KLOE \ combination}, 0.10 < s' <0.95 \text{ GeV}^2) = (489.8 \pm 5.1) \times 10^{-10} ,
\eeq
which is consistent with the individual KLOE measurements and within $1.5\sigma$ of the \allowbreak CMD-2, SND and BESIII measurements, while the difference with the BaBar data is below $3\sigma$. 

\section*{Acknowledgements}

We would like to thank Fedor Ignatov for numerous useful discussions and Daisuke Nomura for his collaboration in producing the compilation and determination of the estimates of $a_{\mu}$. We give special thanks to Maurice Benayoun for his studies and discussions regarding the determination of the pion form factor. We would also like to acknowledge the discussions within the {\em Working Group on Radiative Corrections and MC Generators for Low Energies (Radio MonteCarLOW)} [\texttt{\url{http://www.lnf.infn.it/wg/sighad/}}] and {\em The Muon $(g-2)_{\mu}$ Theory Initiative} concerning this work. The work of Alex Keshavarzi and Thomas Teubner is supported by STFC under the consolidated grants ST/N504130/1 and  ST/L000431/1, respectively.

The KLOE-2 collaboration would like to warmly thank former KLOE colleagues for the access to the data collected during the KLOE data taking campaign.
We thank the DA$\Phi$NE team for their efforts in maintaining low background running conditions and their collaboration during all data taking. We want to thank our technical staff: 
G.F. Fortugno and F. Sborzacchi for their dedication in ensuring efficient operation of the KLOE computing facilities; 
M. Anelli for his continuous attention to the gas system and detector safety; 
A. Balla, M. Gatta, G. Corradi and G. Papalino for electronics maintenance; 
C. Piscitelli for his help during major maintenance periods. 
This work was supported in part 
by the Polish National Science Centre through the Grants No.\
2013/08/M/ST2/00323,
2013/11/B/ST2/04245,
2014/14/E/ST2/00262,
2014/12/S/ST2/00459,
2016/21/N/ST2/01727,
2016/23/N/ST2/ 01293.


\appendix

\section{Properties of a covariance matrix} \label{App:CovMat}

Any covariance matrix, $\mathcal{C}_{ij}$, of dimension $n\times n$ must satisfy the following requirements:
\begin{itemize}
\item As the diagonal elements of any covariance matrix are populated by the corresponding variances, all the diagonal elements of the matrix are positive. Therefore, the trace of the covariance matrix must also be positive
\beq
\text{Trace}(\mathcal{C}_{ij}) = \sum^n_{i=1} \sigma_{ii} = \sum^n_{i=1} \text{Var}_i > 0.
\eeq
\item It is a symmetric matrix, $\mathcal{C}_{ij} = \mathcal{C}_{ji}$, and is, therefore, equal to its transpose, $\mathcal{C}_{ij} = \mathcal{C}_{ij}^{T}$.
\item The covariance matrix is a positive, semi-definite matrix,
\beq {\bf a}^T\mathcal{C} \ {\bf a} \geq 0 \ ; \ {\bf a} \in {\bf R^n},
\eeq where ${\bf a}$ is an eigenvector of the covariance matrix $\mathcal{C}$.
\item Therefore, the corresponding eigenvalues $\lambda_{\bf a}$ of the covariance matrix must be real, greater than or equal to zero and the distinct eigenvectors are orthogonal
\beq
{\bf b} \ \mathcal{C} \ {\bf a} = \lambda_{\bf a}({\bf b}\cdot{\bf a}) = {\bf a} \ \mathcal{C} \ {\bf b} = \lambda_{\bf b}({\bf a}\cdot{\bf b})
\eeq
\beq
\therefore \text{if} \ \lambda_{\bf a} \neq \lambda_{\bf b} \Rightarrow ({\bf a}\cdot{\bf b}) = 0.
\eeq
\item The determinant of the covariance matrix is greater than or equal to zero: $\text{Det}(\mathcal{C}_{ij}) \geq 0$.
\end{itemize}

With error contributions from multiple sources of uncertainty for both statistics and systematics, the contributions of these individual sources must be summed correctly in order to satisfy the necessary conditions for a covariance matrix. In general, should sources of uncertainty be correlated, the element $(i,j)$ of a covariance matrix that describes the total covariance $\sigma_{ij}$ between the two data points should be constructed as 
\beq \label{fullerrorprop}
\mathcal{C}_{ij} \equiv \sigma_{ij} = \sum_{\alpha}\sum_{\beta}\sigma_{i}^{\alpha}\rho^{\alpha\beta}_{ij}\sigma_{j}^{\beta}  \ . \nonumber
\eeq
Here, $\alpha$ and $\beta$ denote individual sources of uncertainty, $\sigma_{i}^{\alpha}$ is the standard deviation of the data point $i$ due to the uncertainty source $\alpha$, $\sigma_{j}^{\beta}$ is the standard deviation of the data point $j$ due to the uncertainty source $\beta$ and $\rho^{\alpha\beta}_{ij}$ is the correlation coefficient that describes the correlation ($-1 \leq \rho \leq 1$) between the uncertainty source $\alpha$ of data point $i$ and the uncertainty source $\beta$ of data point $j$. For the construction of the KLOE covariance matrices, different sources of uncertainty are generally assumed to be independent and, therefore, uncorrelated $(\rho^{\alpha\beta}_{ij}|_{\alpha\neq\beta} = 0)$. Correspondingly, we determine the element $(i,j)$ of the covariances matrices from
\beq \label{KLOEerrorprop}
\mathcal{C}_{ij} = \sum_{\alpha}\rho^{\alpha}_{ij}\sigma_{i}^{\alpha}\sigma_{j}^{\alpha} = \sum_{\alpha}\mathcal{C}^{\alpha}_{ij}  \ ,
\eeq
where $\mathcal{C}^{\alpha}_{ij}$ is the covariance matrix specifically due to the uncertainty source $\alpha$. It follows that to define the total covariance of two data points, we must know the correlation coefficient and absolute error of each data point for each source of uncertainty, which are then summed in accordance with equation~\eqref{KLOEerrorprop}.

\end{document}